\documentclass[letterpaper,twocolumn,10pt]{article}
\usepackage{usenix,epsfig,endnotes}
\usepackage{amsmath}
\usepackage{amssymb}
\usepackage{algorithmic}

\usepackage{array}
\newcolumntype{P}[1]{>{\centering\arraybackslash}p{#1}}
\usepackage{algorithm2e}
\usepackage{enumitem}

\usepackage{booktabs}
\usepackage{tabularx}
\usepackage{subfig}
\usepackage{color, colortbl}
\definecolor{Gray}{gray}{0.9}

\usepackage{stfloats}
\usepackage{float}

\usepackage{graphicx}
\usepackage[table]{xcolor}
\usepackage{appendix}
\graphicspath{{IEEE-SP-Submission/images/}}
\usepackage{subfig}
\usepackage{url}
\usepackage{hyperref}
\usepackage{multirow}
\usepackage[utf8]{inputenc}
\usepackage{color}
\newcommand{\kreminder}[1]{\textcolor{red}{\textbf{kurt: #1}}}
\renewcommand{\kreminder}[1]{{}}
\newcommand{\modificationStart}{\color{red} \bf}
\newcommand{\modificationEnd}{\color{black} \rm  }
\newcommand{\ROneStart}{\color{red}}
\newcommand{\ROneEnd}{\color{black} \rm  }
\newcommand{\response}[1]{\textcolor{blue}{\textbf{(#1)}}}

\begin{document}

%%%%%%%%%%%% FOR REVISION

\onecolumn
\begingroup
\renewcommand{\response}[1]{}
\renewcommand{\modificationStart}{“\color{red}\bf}
\renewcommand{\modificationEnd}{\color{black} \rm” }

\renewcommand{\ROneStart}{``\color{green} \bf}
\renewcommand{\ROneEnd}{\color{black} \rm ''}
\renewcommand{\ROneStart}{``}
\renewcommand{\ROneEnd}{''}

\twocolumn
\endgroup

\setcounter{figure}{0} \renewcommand{\thefigure}{\arabic{figure}}

%%%%%%%%%%%%%%%%%%%%%%%%%%%%%%

%don't want date printed
\date{}

%make title bold and 14 pt font (Latex default is non-bold, 16 pt)
\title{\Large \bf 6thSense: A Context-aware Sensor-based Attack Detector for Smart Devices}

%\def\blind{}
% uncomment next line to show authors
% \undef\blind{}  

\author{
{\rm Amit Kumar Sikder, Hidayet Aksu, A. Selcuk Uluagac}\\
Cyber-Physical Systems Security Lab,\\
Electrical and Computer Engineering Department,\\
Florida International University.\\
\{asikd003, haksu, suluagac\}@fiu.edu\\
}

\maketitle

%\def\ie{{i.e.},~}
%\def\eg{{e.g.},~}
%\def\cf{{c.f.},~}

% Use the following at camera-ready time to suppress page numbers.
% Comment it out when you first submit the paper for review.
\thispagestyle{empty}

\subsection*{Abstract}
Sensors (e.g., light, gyroscope, accelerometer) and sensing enabled applications on a smart device make the applications more user-friendly and efficient. However, the current permission-based sensor management systems of smart devices only focus on certain sensors and any App can get access to other sensors by just accessing the generic sensor API. In this way, attackers can exploit these sensors in numerous ways: they can extract or leak users' sensitive information, transfer malware, or record or steal sensitive information from other nearby devices. In this paper, we propose 6thSense, a context-aware intrusion detection system which enhances the security of smart devices by observing changes in sensor data for different tasks of users and creating a contextual model to distinguish benign and malicious behavior of sensors. %6thSense observes changes in sensor data for different tasks of smartphone users to build a contextual model which is used to distinguish benign and malicious behavior on sensor nodes. 
%We adapt 
6thSense utilizes three different Machine Learning-based detection mechanisms (i.e., Markov Chain, Naive Bayes, and LMT) % Alternative Detection Techniques)
to detect malicious behavior associated with sensors. %%%%% we can say different machine learning based detection mechanisms. 
% 6thSense focuses on utilizing several different Machine Learning-based detection mechanisms from Markov Chain to Naive Bayes to LMT 
%We implemented 6thSense in an Android-based smartphone and tested its efficacy with data collected from real users for various tasks.
We implemented 6thSense on a sensor-rich Android smart device (i.e., smartphone) and collected data from typical daily activities of 50 real users. Furthermore, we evaluated the performance of 6thSense against three sensor-based threats: (1) a malicious App that can be triggered via a sensor (e.g., light), (2) a malicious App that can leak information via a sensor, and (3) a malicious App that can steal data using sensors. Our extensive evaluations show that the 6thSense framework is an effective and practical approach to defeat growing sensor-based threats with an accuracy above 96\% without compromising the normal functionality of the device. Moreover, our framework costs minimal overhead. 
%Did we implement 6thsense? I guess not.

%1. Motivation: why our work is significant\\
%2. problem statement: what is the main problem we are focusing on\\
%3. Approach: How we present our work as a solution to this problem\\
%4. results: What is the outcome of our work\\
%5. Conclusion: What will be the future benefit of this work\\
\section{Introduction}\label{sec:introduction}

Smart devices such as smartphones and smartwatches %wearables 
have become omnipresent in every aspect of human life. Nowadays, the role of smart devices is not limited to making phone calls and messaging only. They are integrated into various applications from home security to health care to military~\cite{chan2009smart, poslad2011ubiquitous}. Since smart devices seamlessly integrate the physical world with the cyber world via their sensors (e.g., light, accelerometer, gyroscope, etc.), they provide more efficient and user-friendly applications~\cite{5560598,s131217292,5606038,Park2011, molaylearning}.  

%On the other hand, 
While the number of applications using different sensors~\cite{lane2011enabling} is increasing and new devices offer more sensors, the presence of sensors have opened novel ways to exploit the smart devices~\cite{6997498}. 
Attackers can exploit the sensors in many different ways \cite{6997498}: they can trigger an existing malware on a device with a simple flashlight~\cite{Hasan:2013:SCH:2484313.2484373}; they can use a sensor (e.g., light sensor) to leak sensitive information; using motion sensors such as accelerometer, and gyroscope, attackers can record or steal sensitive information from other nearby devices (e.g., computers, keyboards) or people~\cite{1301311, Zhuang:2009:KAE:1609956.1609959, Halevi:2012:CLK:2414456.2414509, maiti2015smart}. They can even transfer a specific malware using sensors as a communication channel~\cite{6997498}. Such \textit{sensor-based} threats become more serious with the rapid growth of Apps utilizing many sensors~\cite{norton, ICS-CERT}.

In fact, these sensor-based threats 
highlight %expose 
the flaws of existing sensor management systems %in operating systems 
used by smart devices. Specifically, Android sensor management system relies on permission-based access control, which considers only a few sensors (i.e., microphone, camera, and GPS)\footnote{IOS, Windows, and Blackberry also have permission-based sensor management systems. In this work, we focus on Android.}. Android asks for access permission (i.e., with a list of permissions) only while an App is being installed for the first time. Once this permission is granted, the user has no control over how the listed sensors and other sensors (not listed) will be used by the specific App. Moreover, using some sensors is not considered as a violation of security and privacy in Android. For instance, any App is permitted to access to motion sensors by just accessing the \textit{sensor} API. Access to motion sensors is not controlled in Android.

Existing studies have proposed enhanced access control mechanisms for some of the sensors, but these enhancements do not cover all the sensors of a smart device~\cite{smalley2013security}. Some proposed solutions introduced trusted paths on top of the existing security mechanism for controlling information flow between sensors and Apps, but these are also App-specific solutions and depend upon explicit user consent \cite{jang2014a11y, roesner2012user}. Thus, introducing additional  permission controls for sensors of a smart device will not mitigate the risk of all sensor-based threats as they are App specific and address only data leakage risks. Some  attacks may not abuse sensors directly, instead, they may use sensors as side channels to activate another malware~\cite{joy2011side}. Albeit useful, existing security schemes overlook these critical threats which directly impact the security and privacy of the smart device ecosystem. %Moreover, sensors on smart devices work independently and it is necessary to secure all the different sensors~\cite{antivirus} on a smart device.
Moreover, although sensors on smart devices seem to work independently from each other, a task or activity on a smart device may activate more than one sensor to accomplish the task. Hence, it is necessary to secure all the different sensors~\cite{antivirus} on a smart device and consider the context of the sensors in building any solution against sensor-based threats.

\par In order to address the sensor-based threats, in this paper, we present a novel intrusion detection (IDS) framework called \textit{6thSense}, a comprehensive security solution for sensor-based threats for smart devices. The proposed framework is a \textit{context-aware IDS} and is built upon the observation that for any user activity or task (e.g., texting, making calls, browsing, driving, etc.), a different, but a specific set of sensors becomes active. In a context-aware setting, the 6thSense framework is aware of the sensors activated by each activity or task. %Here, context-aware refers to the ability of inferring user activities by tracking different sensors' data of a device.
6thSense observes sensors data in real time and determines the current use context of the device according to which it concludes whether the current sensor use is malicious or not. 6thSense is context-aware and correlates the sensor data for different user activities (e.g., texting, making calls, browsing, driving, etc.) on the smart device and learns how sensors' data correlates with different activities. As a detection mechanism, 6thSense observes sensors' data and checks against the learned behavior of the sensors. %the ground-truths. 6thSense uses all the sensors to build a context-aware model of a task.  
In 6thSense, the framework utilizes several different %three different 
Machine Learning-based detection mechanisms to catch sensor-based threats including  
%n adapted 
Markov Chain, Naive Bayes, and LMT.  %{\color{red}alternative detection techniques (including other Machine Learning (ML) techniques)}. 
In this paper, we present the design of 6thSense on an Android smartphone because of its large market share~\cite{androidmarket} and its rich set of sensors. To evaluate the efficiency of the framework, we tested it with data collected from real users (50 different users, nine different typical daily activities~\cite{activity}). We also evaluated the performance of 6thSense against three different sensor-based threats and finally analyzed its overhead. Our evaluation shows that 6thSense can detect sensor-based attacks with an accuracy and F-Score over 96\%. Also, our evaluation shows a minimal overhead on the utilization of the system resources.

\textit{\textbf{Contributions:}} In summary, the main contributions of this paper are threefold---
\begin{itemize}
\item \textit{First}, the design of 6thSense, a context-aware IDS to detect sensor-based threats utilizing different machine learning based models from Markov Chain to  Naive Bayes to LMT.   
%three different detection approach: 
%Markov Chain model, Naive Bayes model, and {\color{red}alternative detection techniques}.
%a framework, 6thSense, to detect sensor-based attacks on \textit{Android} smartphone using contextual behavior model. Here, we use two different machine learning algorithms --- Markov Chain Model and Naive Bayes Model to distinguish malicious activities from normal user activities. 
\item \textit{Second}, the extensive performance evaluation of 6thSense with real user experiments over 50 users. 
%to verify our approach, we provide sensor data collected from 35 different users for different activities to train our model and cross-validate with data collected for malicious activities. 
\item \textit{Third}, testing 6thSense against three different sensor-based threats.   
%we provide detail evaluation of our model and present a comparison between both of our approach in the context of detecting sensor-based attacks. We test 6thSense against three different sensor-based attack scenarios and show that 6thSense can work with little overheads and higher accuracy in detecting sensor-based attacks.
\end{itemize}

\textit{\textbf{Organization:}} The rest of the paper is organized as follows: we give an overview of sensor-based threats and existing solutions in Section 2. In section 3, we briefly discuss the \textit{Android's} sensor management system. Adversary model and design facts and assumptions %Different attack scenarios and design facts 
for 6thSense are briefly discussed in Section 4. Different detection techniques used in our framework are described in Section 5. In Sections 6 and 7, we provide a detailed overview of 6thSense including its different components and discuss its effectiveness by analyzing different performance metrics.  Finally, we discuss features and limitations and conclude this paper in Sections 8 and 9, respectively.

\section{Related Work}\label{sec:RelatedWork}
\textit{Sensor-based threats}~\cite{6997498} on mobile devices have become more prevalent than before with the use of different sensors in smartphones such as user's location, keystroke information, etc. Different works \cite{6654855} have investigated the possibility of these threats and presented different potential threats in recent years. One of the most common threats is keystroke inference in smartphones. Smartphones use on-screen QWERTY keyboard which has specific position for each button. When a user types in this keyboard, values in smartphone's motion sensor (i.e., accelerometer and gyroscope) change accordingly~\cite{cai2012practicality}. As different keystrokes yield different, but specific values in motion sensors, typing information on smartphones can be inferred from an unauthorized sensor such as motion sensor data or motion sensor data patterns collected either in the device or from a nearby device can be used to extract users' input in smartphones~\cite{al2013keystrokes, shen2015input, 7113464}. The motion sensor data can be analyzed using different  techniques (e.g., machine learning, frequency domain analysis, shared-memory access, etc.) to improve the accuracy of inference techniques such as  \cite{Aviv:2012:PAS:2420950.2420957, owusu2012accessory, Xu:2012:TIU:2185448.2185465, Miluzzo:2012:TYF:2307636.2307666, Ping:2015:TIL:2766498.2766511,7605458}. Another form of keystroke inference threat can be performed by observing only gyroscope data. Smartphones have a feature of creating vibrations while a user types on the touchpad. The gyroscope is sensitive to this vibrational force and it can be used to distinguish different inputs given by the users on the  touchpad \cite{Narain:2014:SLK:2627393.2627417,Cai:2011:TIK:2028040.2028049, 184479}. Recently, ICS-CERT also issued an alert for accelerometer-based attacks that can deactivate any device by matching vibration frequency of the accelerometer \cite{ICS-CERT, enigma, son2015rocking}. Light sensor readings also change while a user types on the smartphone;  hence, the user input in a smartphone can be inferred by differentiating the light sensor data in normal and typing modes~\cite{Spreitzer:2014:PSE:2666620.2666622}. The light sensor can also be used as a medium to transfer malicious code and trigger message to activate malware~\cite{Hasan:2013:SCH:2484313.2484373,6997498}. The audio sensor of a smartphone can be exploited to launch different malicious attacks (e.g., information leakage, eavesdropping, etc.) on the device. Attackers can infer keystrokes by recording tap noises on touchpad \cite{FooKune:2010:TAP:1866307.1866395}, record conversation of users \cite{schlegel2011soundcomber}, transfer malicious code to the device \cite{6654855, 6997498}, or even replicate voice commands used in voice-enabled different Apps like \textit{Siri}, \textit{Google Voice Search}, etc.~\cite{Diao:2014:YVA:2666620.2666623, 6680832}. Modern smartphone cameras can be used to covertly capture screenshot or video and to infer information about surroundings or user activities \cite{Simon:2013:PSI:2516760.2516770, Meng:2015:CMI:2732198.2732205, Shukla:2014:BYH:2660267.2660360}. GPS of a smartphone can be exploited to perform a false data injection attack on smartphones and infer the location of a specific device~\cite{tippenhauer2011requirements, coffed2014threat}. 

\par \textbf{\textit{Solutions for sensor-based threats:}} Although researchers identified different sensor-based threats in recent years, no complete security mechanism has been proposed that can secure sensors of a smart device. Most of the proposed security mechanisms for smart devices are related to anomaly detection at the application level~\cite{Wang:2015:NHM:2802130.2802132, Sun:2014:DIA:2664243.2664245,Wu:2014:DDA:2663761.2664223,Enck:2014:TIT:2642648.2619091} which are not built with any protection 
%do not provide security 
against sensor-based threats. On the other hand, different methods of intrusion detection have been proposed for wireless sensor networks (WSN)~\cite{strikos2007full,ioannis2007towards,yu2008framework,farooqi2013novel,pongaliur2008securing}, but they are not compatible with smart devices.  
Xu et al. proposed a privacy-aware sensor management framework for smartphones named \textit{Semadroid}~\cite{Xu:2015:SPS:2699026.2699114}, an extension to the existing sensor management system where users could monitor sensor usage of different Apps and invoke different policies to control sensor access by active Apps on a smartphone. Petracca et al. introduced \textit{AuDroid}, a SELinux-based policy framework for smartphones by performing behavior analysis of microphones and speakers~\cite{Petracca:2015:APA:2818000.2818005}. AuDroid controls the flow of information in the audio channel and notifies users whenever an audio channel is requested for access. Jana et al. proposed \textit{DARKLY}, a trust management framework for smartphones which audits applications of different trust levels with different sensor access permissions \cite{jana2013scanner}. Darkly  scans for vulnerability in the source code of an application and try to modify the run-time environment of the device to ensure the privacy of sensor data. \par

\textbf{\textit{Differences from the existing solutions:}} 
Though there is no direct comparable work to compare 6thSense with, differences between existing solutions and our framework can be noted as follows. The main limitation of Semadroid~\cite{Xu:2015:SPS:2699026.2699114} is that the proposed solution is only tested against a similar type of attack scenario (information leakage by a background application). Semadroid also does not provide any extensive performance evaluation for the proposed scheme. %for Semadroid.  
Finally, %Another limitation of this work is that 
this work depends on user permissions to fully enforce an updated policy on the sensor usage which is vulnerable as users might unknowingly approve the sensor permissions for malicious Apps. In another prior work Darkly \cite{jana2013scanner}, the proposed framework is not tested %researchers did not test the framework 
against any sensor-based threats. %which exposes the limitation of this work.
More recent work Audroid presented a policy enforced framework to secure only the audio channels of a smart device. Albeit useful, similar to the others, this work does not consider other sensor-based threats, either. %which is the main shortcoming of this work. 
\textit{Compared to these prior works, 6thSense provides a comprehensive coverage to all the sensors in a smart device and ensures security against three different types of sensor-based threats with high accuracy.}

%%%%%%% I didn't find anything new to add in this section   %%%%%%%%%%%%

\section{Background: Sensor Management in Smart Devices}\label{sec:Background}
\begin{figure}[tb]
  \centering
  \framebox{  
    \includegraphics[height=7cm, width=6cm]{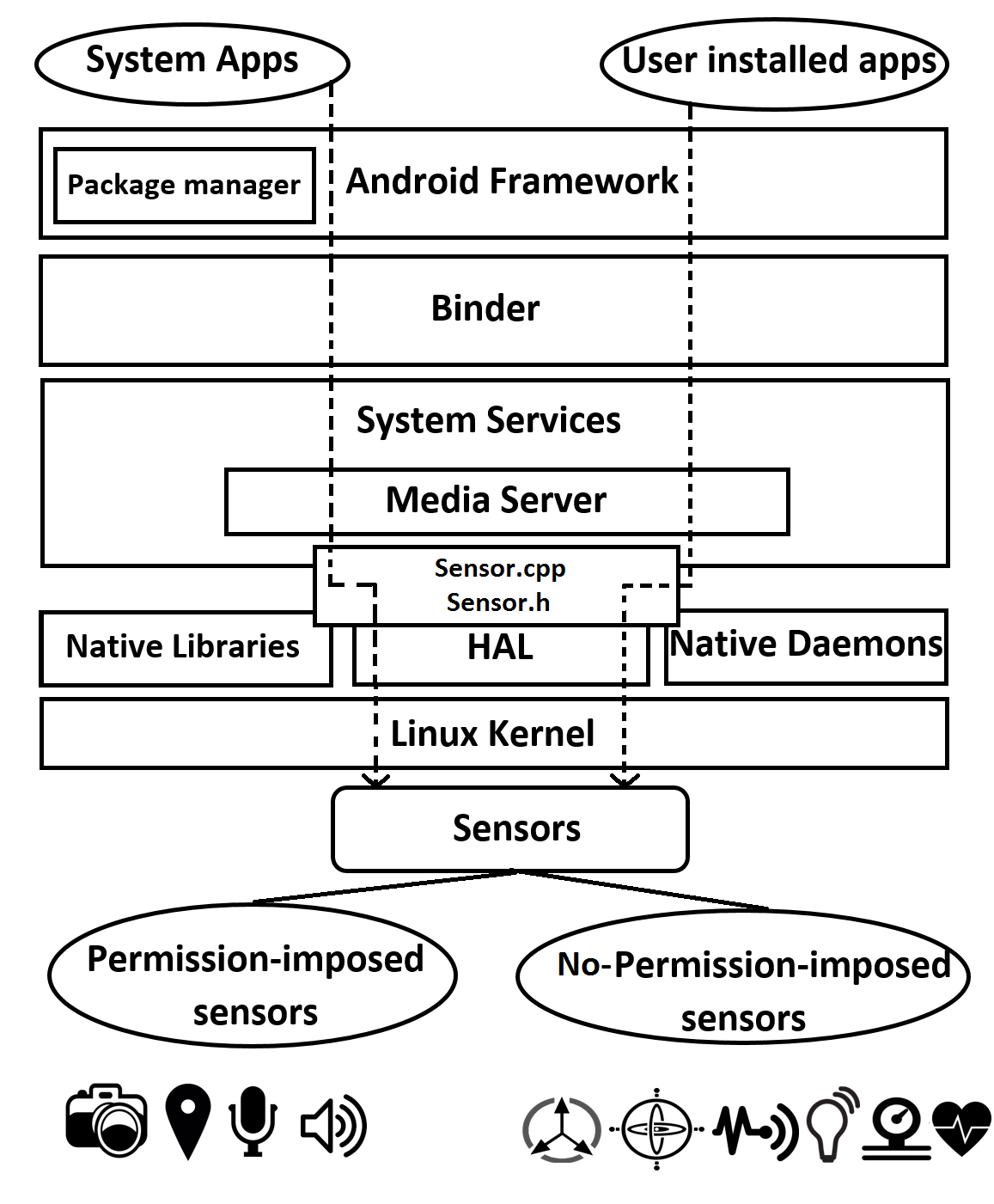}
    }
      \caption{Android Sensor Management Architecture}
      \label{overview}
\end{figure}

Present versions of Android, iOS, or Blackberry do not comprise of any security mechanism to manage the information flow from sensors or among them. %The lack of knowledge about sensor-based threats paves the way for overlooking this design issue. 
For example, any App can get access to motion sensors by just accessing \textit{sensor} API. One task may need more than one sensor, but protecting only one sensor is not a viable design.   %This ignorance about 
The lack of ability to secure the information flow between the sensors and Apps and a holistic view into the utilization of sensors can lead to different malicious scenarios like information leakage, eavesdropping, etc. 
%The Android O/S is widely used in smartphones because of its open-source nature. 
%its open source availability and adaptability in mobile app markets. 

In our work, we focus on Android because of its open-source nature. In Figure~\ref{overview}, we present how Android handles access to different sensors by Apps (installed by the user) and system Apps (installed automatically by Android). Apps access to sensors by sending requests via Software Development Kit (SDK) API platform which then registers the App to a corresponding sensor~\cite{milette2012professional}. If more than one App tries to access the same sensor, the SDK API runs a multiplexing process which enables different Apps to be registered in the same sensor. Hardware Abstraction Layer (HAL) works as an interface to bind the sensor hardware with the device drivers in Android. HAL has two parts: \textit{Sensors.h} works as HAL interface and \textit{Sensors.cpp} works as the HAL implementation. Through the HAL library, different applications can communicate with the underlying Linux kernel to read and write files associated with sensors. For most of the sensors, no permission is needed to access these files. For \textit{permission-imposed sensors} (i.e., camera, microphone, and GPS), a permission is explicitly needed from the user to ensure file access to a specific App. This user permission is declared inside the \textit{AndroidManifest.xml} file of an App and once the user accepts the permission, that App can have access to the corresponding sensor and other \textit{no-permission imposed sensors} even without any explicit approval from the users. This lack of security in sensor access can lead to different malicious attacks on a device. 
% One possible solution to minimize this design flaw is to lock down sensor API with permission. However, locking down sensor API with permissionin OS level will not surpass sensor-based threats as users are not aware of these threats yet and can allow malicious Apps to use sensors unknowingly.
% here is a question why would not a simple extra control on the no-permission imposed sensor solve this problem? 

\section{Adversary Model and Assumptions}\label{sec:Adversarymodel}
In this section, we discuss different threats that may use sensors to execute malicious activities on a smart device. Different design assumptions are also explained in this section.
\subsection{Adversary Model}
%In the previous section, we already discussed existing malicious attacks on mobile devices. 
For this work, we consider the following sensor-based threats similar to \cite{6997498}: 
\begin{itemize}
\item \textbf{\textit{Threat 1}-Triggering a malicious App via a sensor.} A malicious App can exist in the smart device which can be triggered by sending a specific sensory pattern or message via sensors. %For our experiment, we developed two different apps which can be triggered using the light sensor and motion sensors on the smartphone. 
\item \textbf{\textit{Threat 2}-Information leakage via a sensor.} A malicious App can exist in the device which can leak information to any third party using sensors. %For our experiment, we developed a malware that could record conversations as audio clips and playback after a specific time to leak information. This attack scenario included both the microphone and speaker on the smartphone.
\item \textbf{\textit{Threat 3}-Stealing information via a sensor.} %Triggering a sensor
A malicious App can exist in the device which can exploit the sensors of a smart device and start stealing information after inferring a specific device mode (e.g., sleeping). %For our study, we developed a malicious app that could scan all the sensors and if none of the sensors were changing their state, the malicious could open up the camera and record videos surreptitiously.
\end{itemize}

In this paper, we cover these three types of malicious sensor-based threats.  %associated with the sensors. 
We also note that to build our adversary model, we consider any component on a smart device that interacts with the physical world as a sensor \cite{Petracca:2015:APA:2818000.2818005}. In section 7, we show how 6thSense defends against these threats.  %Attack scenario 1 represents triggering a malware on a device. On the other hand, attack scenario 2 and 3 represent both eavesdropping and information leakage. 
\subsection{Design Assumptions and Features}
%Overview of Our Approach}
%For developing a comprehensive security scheme for sensors, several design issues need to be addressed correctly. 
In designing a comprehensive security scheme like 6thSense for sensor-based threats, we note the following design assumptions and features:
\begin{itemize}
\item \textbf{\textit{Sensor co-dependence:}} A sensor in a smart device is normally considered as an independent entity on the device. Thus, one sensor does not know what is happening in another sensor. However, %To address sensor dependency, 
in this work, we consider sensors as co-dependent entities on a device instead of independent entities. The reason for this stems from the fact that for each user activity or task on a smart device, a specific set of sensors remains active. For example, if a user is walking with a phone in hand, motion sensors (i.e., gyroscope, accelerometer), the light sensor, GPS will be active. On the contrary, if the user is walking with the phone in the pocket or bag,  instead of the light sensor, the proximity sensor will remain active. Thus, a co-dependent relationship exists  between sensors while performing a specific task. Each activity uses different, but specific set of sensors to perform the task efficiently. Hence, one can distinguish the user activity by observing the \textit{context of the sensors} for a specific task. 6thSense uses the context of all the sensors to distinguish between normal user activities and malicious activities. In summary, %one can state that 
\textit{sensors in a smart device are individually independent, but per activity-wise dependent} and 6thSense considers the context of the activities in its design. 

\item \textbf{\textit{Adaptive sensor sampling:}} Different sensors have different sampling frequencies. To monitor all the sensor data for a specific time, a developed solution must consider and sample the sensor data correctly. %If an individual sensor frequency is considered for sampling each sensor data, there will be missing data for the same time instance which will introduce error in the  detection technique. We address this design issue in the design of 6thSense. 
Our proposed framework considers sampling the sensor data over a certain time period instead of individual sensor frequencies 
%sampled dataset of all the sensors over a certain time period 
which mitigates any possible error in processing of data from different sensors.
\item \textbf{\textit{Faster computation:}} Modern high precision sensors on smart devices have high resolution and sampling rate. As a result, sensors provide large volume of data even for a small time interval. %Modern sensors have high resolution and higher sampling rate. 
A solution for sensor-based threats should quickly process any large data  from different sensors in real time while ensuring a high detection rate. To address this, %issue 
we use different machine learning algorithms which are proven simple and fast techniques \cite{aviles2011comparison, sahs2012machine}.
\item \textbf{\textit{Real-time monitoring:}} 6thSense provides real-time monitoring to all the sensors which mitigates the possibility of data tempering or false data injection on the device. 
\end{itemize}

%Thus, 6thSense is designed to cope with all the threats mentioned above.
%To overcome these challenges, we consider a context-aware IDS model for 6thSense. Our main assumption is that for each user activity on smartphone, a specific set of sensors remain active. Thus, rather than considering each sensor as an independent entity, we consider sensors as logically co-dependent entities while performing a specific task. This co-dependent relationship between sensors can be seen in all the activities on the smartphone.  Considering a sensor as co-dependent entity overcomes the first challenge mentioned above.  On top of context-aware scheme, 6thSense considers sampling the sensor data over a certain time period instead of individual sensor frequencies 
%sampled dataset of all the sensors over a certain time period 
%which mitigates errors in data processing of different sensors.% at the same time. 
%Finally, we are using three approaches - Markov Chain, Naive Bayes model, and machine learning tool \textit{WEKA}  
%\section{Analytical Model}\label{sec:AnalyticalModel}
\section{Detection Techniques: Theoretical Foundation}\label{sec:AnalyticalModel}
In this section, we describe the details of the detection techniques  used in 6thSense from a theoretical perspective. 

%in context of the sensory channels of a smartphone. In recent days, machine learning algorithms gain much popularity in developing intrusion detection systems for their easy deployment architecture and low false alarm rate. Machine learning also offers different mathematical model which offers developers to choose the best fit for their IDS. 

For the context-aware IDS in 6thSense, we utilize several different machine learning-based techniques including Markov Chain~\cite{brooks2011handbook}, Naive Bayes~\cite{murphy2006naive} and alternative set of ML algorithms (e.g., PART, Logistic Function, J48, LMT, Hoeffding Tree, and Multilayer Perception) 
%three differentapproaches (Markov Chain~\cite{brooks2011handbook}, Naive Bayes model~\cite{murphy2006naive}, and alternative detection techniques) 
to differentiate between normal behavior from malicious behavior on a smart device. The main advantage of using Markov Chain model is that it is easy to build the model from a large dataset and computational requirements are modest which can be met by resource-limited devices. As smart devices have less processing speed, a Markov Chain-based approach can work smoothly in the context of sensor data analysis. On the other hand, Naive Bayes technique is chosen for its fast computation rate, small training dataset requirement, and ability to modify it with new training data without rebuilding the model from scratch. Other ML techniques are also common in malware detection because of higher accuracy rate. A brief discussion of these approaches in the context of 6thSense is given below. The efficacy of these different approaches utilized in 6thSense is analyzed in Section 7.

\subsection{Markov Chain-Based Detection}
A Markov Chain-based detection model can be described as a discrete-time stochastic process which denotes a set of random variables and defines how these variables change over time. 
%\begin{itemize}
%\item Probability distribution of the state at time \textit{t+1} depends on the state at time \textit{t} only. There is no effect of previous states leading to the state at time \textit{t} over probability distribution at time \textit{t+1}. Here, the state refers to the overall condition of the stochastic process.
%\item A state transition from previous timestamp (\textit{t}) to next timestamp (\textit{t+1}) is independent of time.
%\end{itemize}
Markov Chain can be applied to illustrate a series of events where and what state will occur next depends only on the previous state. In 6thSense, a series of events represents user activity and state represents sensor conditions (i.e., sensor values, on/off status) of the sensors in a smart device. We can represent the probabilistic condition of Markov Chain as in Equation 1 where \textit{$X_t$} denotes the state at time \textit{t} \cite{keilson2012markov}: 

\begin{equation}
\begin{split}
\begin{aligned}
P(X_{t+1} = x| X_1 = x_1, X_2 = x_2 ..., X_t = x_t) = 
\\P(X_{t+1} = x| X_t = X_t) , \\ 
when,\ P(X_1 = x_1, X_2 = x_2 ..., X_t = x_t) > 0
\end{aligned}
\end{split}
\end{equation}

In 6thSense, we observe the changes of the  conditions of a set of sensors as a variable which changes over time. The condition of a sensor indicates whether the sensor value is changing or not from a previous sensor value in time. As such, \textit{S} denotes a set which represents current conditions of \textit{n} number of sensors. So, \textit{S} can be represented as follows.

\begin{equation}
\begin{aligned}
S = \{S_1, S_2, S_3, ... , S_n\}, 
\\ S_1, S_2, S_3, ... , S_n = 0\ or\ 1
\end{aligned}
\end{equation}

For 6thSense, we use a modified version of the general Markov Chain. Here, instead of predicting the next state, 6thSense determines the probability of a transition occurring between two states at a given time. In 6thSense, the Markov Chain model is trained  with a training dataset collected from real users and the transition matrix is built  accordingly. Then, 6thSense determines conditions of sensors for time \textit{t} and \textit{t+1}. Let us assume, \textit{a} and \textit{b} are a sensor's state in time \textit{t} and \textit{t+1}. 6thSense looks up for the probability of transition from state \textit{a} to \textit{b} which can be found by looking up in the transition matrix, \textit{P} and calculating \textit{P(a,b)}. As the training dataset consists sensor data from benign activities, we can assume that, if transition from state \textit{a} to \textit{b} is malicious, the calculated probability from transition matrix will be zero. Details of this % adaptation of 
Markov Chain-based detection model in 6thSense are given in Appendix A1. 

\subsection{Naive Bayes Based Detection}
Naive Bayes model is a simple probability estimation method which is based on Bayes' method. The main assumption of the Naive Bayes detection is that the presence of a particular sensor condition in a task/activity has no influence over the presence of any other feature on that particular event. The probability of each event can be calculated by observing the presence of a set of specific features.

%Assume, \textit{${p(x_1,x_2)}$} is the general probability distribution of two events \textit{${x_1, x_2}$}. Using the Bayes rule, we can have the following equation: 
%\begin{equation}
%p(x_1,x_2) = p(x_1|x_2)p(x_2)
%\end{equation}
%where, \textit{${p(x_1|x_2)}$} = Probability of event \textit{${x_1}$} given that event \textit{${x_2}$} will happen. Now, if we have another variable, \textit{c}, we can rewrite Equation 7 as follows: 
%\begin{equation}
%p(x_1,x_2|c) = p(x_1|x_2,c)p(x_2|c)
%\end{equation}
%If knowledge of \textit{c} is sufficient enough to determine the probability of event \textit{${x_1}$}, we can state that there is conditional independence between \textit{${x_1}$} and \textit{${x_2}$}~\cite{panda2007network}. So, we can rewrite the first part of Equation 8 as \textit{${p(x_1|x_2,c) = p(x_1|c)}$}, which modifies Equation 8 as follows: 
%\begin{equation}
%p(x_1,x_2|c) = p(x_1|c)p(x_2|c)
%\end{equation}

6thSense considers users' activity as a combination of \textit{n} number of sensors. Assume X is a set which represents current conditions of \textit{n} number of sensors. We consider that conditions of sensors are conditionally independent (See Section 4.2), which means a change in one sensor's working condition (i.e., on/off states) has no effect over a change in another sensor's working condition. As explained earlier, the probability of executing a task depends on the conditions of a specific set of sensors. So, in summary, although one sensors' condition does not control another sensor's condition, overall the probability of executing a specific task depends on all the sensors' conditions. As an example, if a person is walking with his smartphone in his hand, the motion sensors (accelerometer and gyroscope) will change. However, this change will not force the light sensor or the proximity sensor to change its condition. Thus, sensors in a smartphone change their conditions independently, but execute a task together. We can have a generalized model for this context-aware detection \cite{mukherjee2012intrusion} as follows: 

\begin{equation}
p(X|c)=\prod_{i=1}^{n} p(X_i|c)
\label{}
\end{equation}
Detailed description of this Naive Bayes model in 6thSense is given in Appendix A2.

\subsection{Alternative Detection Techniques}
In addition to  Markov Chain and Naive Bayes models above, there are other machine learning algorithms (such as  PART, Logistic Function, J48, LMT, Hoeffding Tree, and Multilayer Perception) that are very popular for anomaly detection frameworks because of their faster computation ability and easy implementation feature. 
%WEKA is a data mining tool which offers data analysis using different machine learning approaches \cite{schmidt2009static}. Basically, WEKA is a collection of machine learning algorithms developed at the University of Waikato, New Zealand, which can be directly applied to a dataset or can be integrated with a framework using JAVA platform \cite{peiravian2013machine}. WEKA offers different types of classifier to analyze and build predictive model from given dataset. \par
In the alternative detection techniques, we used four types of ML-based classifier to build an analytical model for 6thSense. The following briefly discusses these classifiers and our rationale to include them. 

\par \textit{Rule-based Learning.} Rule-based ML works by identifying a set of relational rules between attributes of a given dataset and represents the model observed by the system~\cite{gu2007bothunter}. The main advantage of the rule-based learning is that it identifies a single model which can be applied commonly to any instances of the dataset to make a prediction of outcome. As we train 6thSense with different user activities, the rule-based learning provides one model to predict data for all the user activities which simplifies the framework. For 6thSense, we chose, \textit{PART} algorithm for the rule-based learning. 

\par\textit{Regression Model.} Regression model is widely used in data mining for its faster computation ability. This type of classifier observes the relations between dependent and independent variables to build a prediction model \cite{dahl2013large, wu2012droidmat}. For 6thSense, we have a total 11 attributes where we have one dependent variable (device state: malicious/benign) and ten independent variables (sensor conditions). Regression model observes the change in the dependent variable by changing the values of the independent variables and build the prediction model. We use the logistic regression model in 6thSense,  which performs with high accuracy against conventional Android malware~\cite{shabtai2012andromaly}. 

\par \textit{Neural Network.} Neural network is another common technique that is being adapted by researchers for malware detection. In neural network techniques, the relation between attributes of dataset is compared with the biological neurons and a relation map is created to observe the changes for each attribute~\cite{linda2009neural}. We chose \textit{Multilayer Perceptron} algorithm for training the 6thSense framework as it can distinguish relationships among  non-linear dataset. 

\par \textit{Decision Tree.} Decision tree algorithms are predictive models where decision maps are created by observing the changes in one attribute in different instances \cite{ye2007imds}. These types of algorithms are mostly used in a prediction model where output can have a finite set of values. For 6thSense, we utilized and tested three different decision tree algorithms (\textit{J48}, \textit{LMT (Logistic Model Tree)}, and \textit{Hoeffding tree}) to compare the outcome of our framework.
%To implement ML-based detection in 6thSense, we used WEKA, a data mining tool which offers data analysis using different machine learning approaches \cite{schmidt2009static}. Basically, WEKA is a collection of machine learning algorithms developed at the University of Waikato, New Zealand, which can be directly applied to a dataset or can be integrated with a framework using JAVA platform \cite{peiravian2013machine}. WEKA offers different types of classifier to analyze and build predictive model from given dataset. \par
\section{6thSense Framework} \label{sec:Framework}
\begin{figure}[tb]
  \centering
  \framebox{
    \includegraphics[height=7cm, width=6cm]{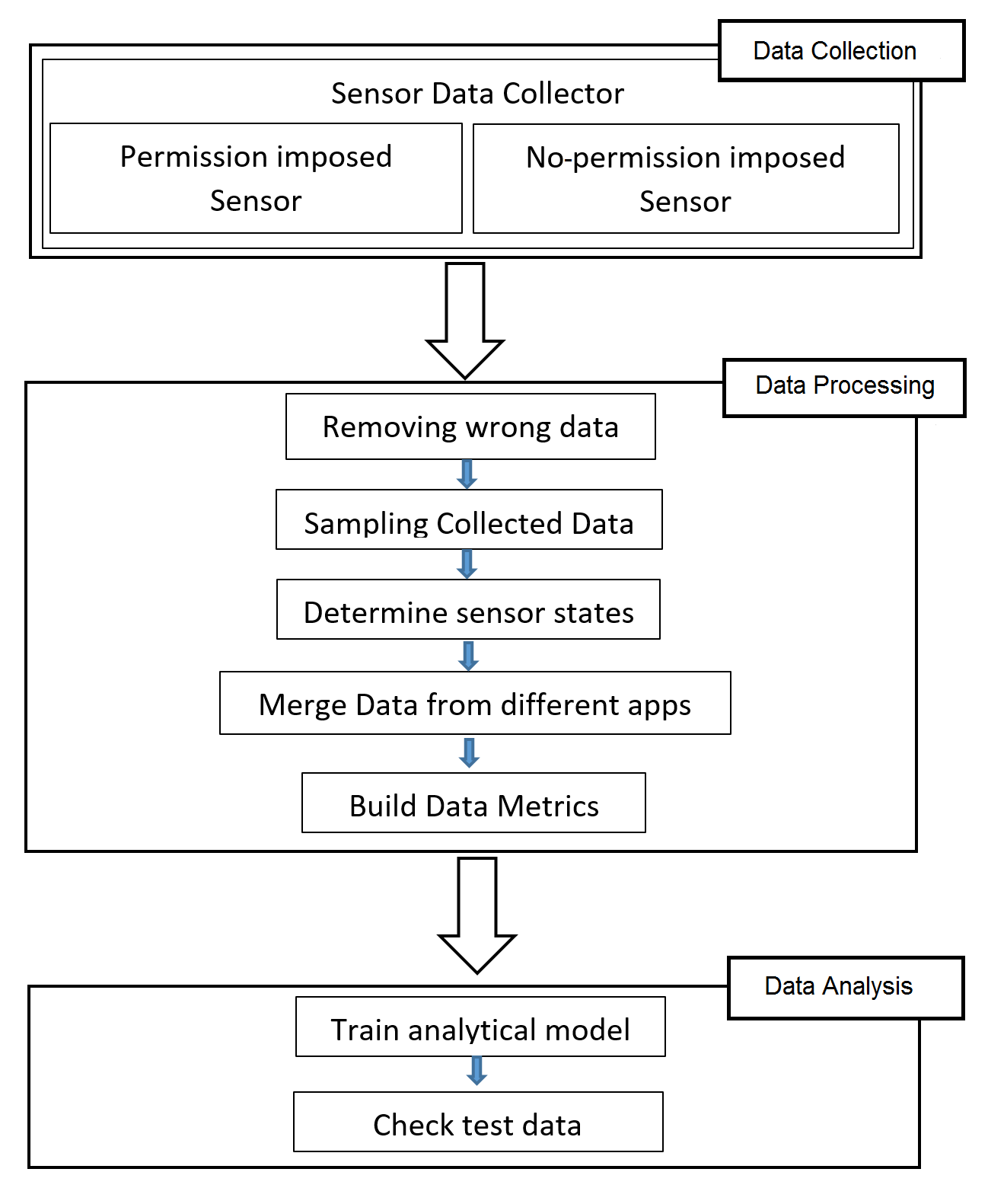}
    }
      \caption{Overview of 6thSense.}
      \label{overviewfig}
\end{figure}

% \vspace{-1cm}
In this section, we provide a detailed overview of our proposed contextual behavior IDS framework, 6thSense, for detecting sensor-based threats on smart devices. As illustrated in Figure~\ref{overviewfig}, 6thSense has three main phases: (1) data collection, (2) data processing, and (3) data analysis. In the data collection phase, we use a custom Android application %two different applications 
to collect the sensor data for different user activities and 
the collected sensor data are then processed in the data processing phase.  %We use two different applications as 
Note that in 6thSense some sensors provide discrete values as data (e.g., accelerometer, gyroscope, light sensor, etc.) while other sensors provide their on-off state as sensor data (e.g., microphone, speaker, etc.).  %To avoid data mismatch and error in data sampling, we create two different apps for different types of sensors. 
In phase 3, the collected data is fed into detection models and the end result indicates whether the current state of the device is malicious or not. The following sub-sections briefly describe these three phases.
% in  to the order.

\subsection{Data Collection Phase}
In this phase, 6thSense collects data from different sensors of a smart device. There can be multiple sensors in a smart device. We chose nine sensors in total to identify different user activities using a sensor-rich Android device. The sensors  selected are accelerometer, gyroscope, light sensor, proximity sensor, GPS, audio sensor (microphone and speaker), camera, and headphone. 6thSense does not consider all the other sensors available in the device because all typical user activities do not affect all the sensor values. For example, the gravity sensor value does not change effectively while talking or walking with the phone. The chosen sensors are then categorized into two following categories.

\begin{itemize}
\item \textit{No-permission-imposed sensors:}  No-permission-imposed sensors can be defined as sensors that do not need any user permission explicitly to be  accessed by an App. For 6thSense, we chose four no-permission imposed sensors (i.e., accelerometer, gyroscope, light, proximity sensors). We can also refer these sensors as data-oriented sensors in the context of 6thSense because values provided by these sensors need to be observed to infer user activities. For example, accelerometer's and gyroscope's values change with motion and they give values on \textit{X, Y,} and \textit{Z} axes. These values change along with the motion in different axes. To detect whether a sensor is activated or not for a specific activity, one needs to observe values of these sensors. 

\item \textit{Permission-imposed sensors:} Permission-imposed sensors are those which need user permission to be accessed by an App. For 6thSense, we chose five permission-imposed sensors to build the context-aware model (camera, microphone, GPS, speaker, and headset). The conditions of these sensors can be represented by their logical states (on/off status) for different user activities. Hence, we also referred to these sensors as logic-oriented sensors in the context of 6thSense. For example, camera has only two values to identify users' activity: on and off. So, it can be represented with 0 or 1 to detect if the camera is on or off correspondingly.  
\end{itemize}

% Accelerometer, gyroscope, light sensor, and proximity sensor are categorized as data-oriented sensors while all others are categorized as logic-oriented sensor.

\par To collect the data and logical values from sensors, we built a custom Android App and 6thSense used this in the data collection phase.  
%two apps named - sensor value collector and sensor logic state detector. \textit{Sensor value collector} collects numerical values of data oriented sensors.
In Android, this App uses \textit{sensoreventlistener} API to log numerical values of the data-oriented sensors. On the other hand, the App  
%\textit{sensor logic state detector} 
determines the state of the sensor and logs 0 or 1 if the sensor is on or off, respectively. This App uses the user permission access to use the microphone, GPS, and camera to record the working condition of these sensors. For GPS, we consider two datasets - either GPS is turned on or not and either location is changing or not. In total, six different logic state information for five aforementioned permission-imposed sensors are  collected by this App.
%\begin{figure*}[!ht]
% \centering
%\subfloat [Sensor Value Collector]{ \framebox{
  
    % \includegraphics[width=\textwidth,height=7cm]{new3-1.png}
%   \includegraphics[width=7cm,height=8cm]{flow}
%    }
%}
%\subfloat [Sensor Logic State Detector]{ \framebox{
  
    % \includegraphics[width=\textwidth,height=7cm]{new3-1.png}
%    \includegraphics[width=7cm,height=8cm]{flow_2}
%    }
%}
%      \caption{Flow chart of data collection apps}
%      \label{flow}
%\end{figure*}
%\begin{figure}[!ht]
%  \centering
%    \includegraphics[height=10cm, width=20cm]{flow}
%      \caption{Flow Chart of Android data collection app}
%\end{figure}

Note that we chose different typical daily human activities~\cite{activity4} that involve the smart device to build our contextual model. These activities include walking (with phone in hand and pocket), talking, interacting (playing games, browsing, listening to music), video calling, driving (as driver and passenger). Furthermore, the number of  activities is configurable in 6thSense and is not limited to aforementioned examples. In the evaluation of 6thSense, we chose a total of nine typical daily activities as they are considered as common user activities for a smart device~\cite{activity4}. We collect these data using the App for different users to train the 6thSense framework which is then used to distinguish the normal sensor behavior from the malicious behavior. 
In summary, the aforementioned App collects data from nine different sensors for nine typical user activities. We observe sensor state (combination of working conditions (i.e., values, on/off status) of nine different sensors) in a per second manner for each user activity. Each second of data for user activity corresponds to 1024 state information from nine different sensors. 
\begin{table*}[htb!]
\centering
\fontsize{8}{10}\selectfont
\begin{tabular}{cccc}
\toprule
\textbf{\begin{tabular}[c]{@{}c@{}}Sensor \\ type\end{tabular}} & \textbf{Name} & \textbf{Model} & \textbf{\begin{tabular}[c]{@{}c@{}}Specification\end{tabular}} \\ \hline
\midrule
\multirow{4}{*}{\begin{tabular}[c]{@{}c@{}}No-permission imposed \\sensors\end{tabular}}  & Accelerometer  & MPU6500 Acceleration Sensor   & 19.6133 $m/s^2$, 203.60 Hz, 0.25 mA \\ \cline{2-4} 
                         & Gyroscope      & MPU6500 Gyroscope Sensor    & 8.726646 rad/s, 203.60 Hz, 6.1 mA \\\cline{2-4} 
                         & Light Sensor   & TMG399X RGB Sensor      	& 600000 lux, 5.62 Hz, 0.75 mA   \\\cline{2-4}
                         & Proximity Sensor  & TMG399X proximity sensor &  8V, 0.75 mA   \\\cline{2-4}
                         \hline
\multirow{3}{*}{\begin{tabular}[c]{@{}c@{}}Permission-imposed sensors\end{tabular}} & Camera      & Samsung S5K2P2XX    & 12 megapixels, 30 fps, 4.7 mA        \\ \cline{2-4} 
                          & Microphone  &  \begin{tabular}[c]{@{}c@{}}Qualcomm Snapdragon \\801 Processor built in microphone\end{tabular}   & 86 dB, .75 mA          \\ \cline{2-4} 
                          & Speaker     &  \begin{tabular}[c]{@{}c@{}}Qualcomm Snapdragon \\801 Processor built in speaker\end{tabular}   & 110 dB, 1 mA                                                                 \\ \hline
\bottomrule                                                                          
\end{tabular}
\caption{Sensor list of Samsung Galaxy S5 Duo used in experiment.}
\label{tab:sensors}
\end{table*}

\subsection{Data Processing Phase}
After the data collection, in the second phase of the framework, we organize the data to use in the proposed IDS framework. As different sensors have different frequencies on the smart device, the total number of readings of sensors for a specific time period is different. %We have to sample data to use them for analysis part. 
For example, the accelerometer and gyroscope of \textit{Samsung Galaxy S5} have a sampling frequency of approximately 202 Hz while the light sensor has a sampling frequency of 5.62 Hz. Thus, the data collected in Phase 1 needs to be sampled and reorganized. 6thSense observes the change in the sensor condition in each second to determine the overall state of our device and from this per second change, 6thSense determines the activity of users. For this reason, 6thSense takes all the data given by a single sensor in a second and calculates the average value of the sensor reading. This process is only applicable for the data oriented sensors as mentioned earlier. Again, the data collected from the App  %\textit{sensor value collector}
is numerical value given by the  sensor. However, for the  detection %analytical 
model, we only consider the condition of the sensors. 6thSense observes the data collected by the aforementioned App and determines whether the condition of sensors is changing or not. If the sensor value is changing from the previous value in time, 6thSense represents the sensor condition as 1 and 0 otherwise. 
% We use \textit{MATLAB} for data processing. 
%The data collected from the \textit{sensor logic state detector} 
The logic state information collected from the sensors 
need to be reorganized, too as these data are merged with the data collected from the collected values from the other sensors  %\textit{sensor value collector} 
to create an input matrix. The sampling frequency of the logical state detection is   %\textit{sensor logic state detector} app is 
0.2 Hz which means in every five seconds the App generates one session of dataset. We consider the condition of the sensors to be the same over this time period and organize the data accordingly. The reorganized data generated from the aforementioned App are then merged to create the training matrices.

\subsection{Data Analysis Phase}
In the third and final phase,  6thSense uses different machine learning-based detection techniques introduced in the  previous section to analyze the data matrices generated in the previous phase. 
% We use \textit{MATLAB} to create two different scripts and generate two set of results for same test data matrix. 

\par For the Markov Chain-based detection, we use 75\% of the collected data to train 6thSense and generate the transition matrix. This transition matrix is used to determine whether the transition from one state to another is appropriate or not. Here, state refers to generic representation of all the sensors' conditions on a device. For testing purposes we have two different data set --- basic activities or trusted model and malicious activities or threat model. The trusted model consists of 25\% of the collected data for different user activities. We test the trusted model to ensure the accuracy of the 6thSense framework in detecting benign activities. The threat model is built from performing the attack scenarios mentioned in Section~\ref{sec:Adversarymodel}. We calculate the probability of a transition occurring between two states at a given time and accumulate the total probability to distinguish between normal and malicious activities. 

\par To implement the Naive Bayes-based detection technique, we use the training sessions to define different user activities. In 6thSense, we have nine typical user activities in total as listed in Table~\ref{tasklist}. We use groundtruth user data to define these activities. Using the theoretical foundation explained in Section~\ref{sec:AnalyticalModel}, we calculate the probability of a test session to belong to any of these defined activities. As we consider one second of data in each computational cycle, we calculate the total probability up to a predefined configurable time interval (in this case five minutes). This calculated probability is used to detect malicious activities from normal activities. If the computed probability for all the known benign activities is not over a predefined threshold, then it is detected as a malicious activity. 

\par For the other alternative machine-learning-based detection techniques, we used WEKA, a data mining tool which offers data analysis using different machine learning approaches \cite{schmidt2009static, hall2009weka}. Basically, WEKA is a collection of machine learning algorithms developed at the University of Waikato, New Zealand, which can be directly applied to a dataset or can be integrated with a framework using JAVA platform \cite{peiravian2013machine}. WEKA offers different types of classifier to analyze and build predictive model from given dataset. We use 10 fold cross-validation method to train and test 6thSense with different ML techniques in Section 7. 

\section{Performance Evaluation of 6thSense} \label{sec:Evaluation}
In this section, we evaluate the efficiency of the proposed context-aware IDS framework, 6thSense, in detecting the sensor-based threats on a smart device. We test 6thSense with the data collected from different users for benign activities and adversary model  described in Section~\ref{sec:Adversarymodel}. As discussed earlier, 6thSense  considers three sensor-based threats: (1) a malicious App that can be triggered via a light or motion sensors, (2) a malicious App that can leak information via audio sensor, and (3) a malicious App that steals data via camera. %sensor.  % 
Furthermore, we measured the performance impact of 6thSense on the device %with different variables 
and present a detailed results for the efficiency of the 6thSense framework. Finally, we discuss the performance overhead of the framework in this section.
%% haksu: Amit please look at meaning of efficiency and effectiveness. 
% Efficiency seems the correct word. Other IDS papers used efficiency. Effectiveness only focus on how correctly it is detecting attacks.  

\subsection{Training Environment} \label{sec:trainingenv}
In order to test the effectiveness of 6thSense, we implemented it on a sensor-rich Android-based smartphone. However, our framework would also efficiently work in another smart device such as smartwatch. In the evaluations, we used \textit{Samsung Galaxy S5 Duos} as a reference Android device to collect sensor data for different typical user activities. We chose this Android device as \textit{Samsung} currently holds approximately 20.7\% of total marketshare of smartphone~\cite{Samsung} and provides a rich set of sensors. A list of sensors of \textit{Samsung Galaxy S5 Duos} is given in Table~\ref{tab:sensors}. As discussed earlier, we  selected 9 different typical user activities or tasks to collect user data. These are typical basic activities with smartphones that people usually do in their daily lives~\cite{activity}. The user activities/tasks are categorized in two categories as generic activities and user related activities.

\par \textit{Generic activities} are the activities in which the sensor readings are not affected by the smartphone users.  Sleeping, driving with the phone using GPS as a navigator, and driving with phone in pocket are three generic activities that we considered in this work.  Basically, in the generic activities, sensors' data are not affected by different users since the smart phone is not in contact with the user or user is not directly interacting with the phone. For \textit{user- related activities}, in which the sensor readings may be affected by the device user, we identified six different activities including walking with the phone in hand, playing games, browsing, and making voice calls and video calls.

%\begin{figure*}[htbp]
%\centering
%\subfloat[False Negative and True Negative Rate]{\includegraphics[height=4.8cm,width=5.5cm]{Markov_false_and_true_positive}}
%\subfloat[Accuracy and F-score]{\includegraphics[height=4.8cm, width=5.5cm]{Accuracy_vs_F-score_markov}}
%\subfloat[ROC Curve]{\includegraphics[height=4.8cm, width=5.5cm]{ROC_markov}}
%\caption{Performance evaluation of Markov Chain based approach}
%\label{markovfig}
%\end{figure*}
6thSense was tested by 50 different individuals aged from 18 to 45 while the sensor data was collected. We note that our study with human subjects was approved by the appropriate Institutional Review Board (IRB) and we followed all the procedures strictly in our study. Each participant received some monetary compensation for participating in our experiments. To ensure privacy and anonymity, we used fake user IDs rather than any personal information. We collected 300 sets of data for six user-related activities where each dataset comprised of 5 minutes long data from the selected nine sensors mentioned in Section~\ref{sec:Framework}. We also collected three sets of data for each general activity. We asked the different users to perform the same activity to ensure the integrity for different tasks. Note that each five minute of data collected for user related and generic activities corresponds to 300 events with 1024 different states. Here, states represent a combination of conditions (i.e., values, on/off status) of nine different sensors and events represent user activities per second. So, a total of 307,200 different event-state information were  analyzed by 6thSense.

\par For the malicious dataset, we created three different attack scenarios considering the adversary model  mentioned in Section 4. For Threat 1, we
developed two different Android Apps which could be triggered using the light sensor and motion sensors on the smartphone. To perform the attack described in Threat 2, we developed a malware that could record conversations as audio clips and playback after a specific time to leak the information. This attack scenario included both the microphone and speaker on the smartphone. For Threat 3, we developed a malicious App that could scan all the sensors and if none of the sensors were changing their working
\begin{table}[h!]
\centering
\begin{tabular}{p{3cm}p{3.7cm}}
\toprule
\textbf{Task Category}                            & \textbf{Task Name}   \\ \hline
\midrule
\multirow{3}{*}{Generic Activities}      & 1. Sleeping                         \\ \cline{2-2} 
                                         & 2. Driving as driver                \\ \cline{2-2} 
                                         & 3. Driving as passenger             \\ \hline
\multirow{7}{*}{User-related Activities} & 1. Walking with phone in hand       \\ \cline{2-2} 
                                         & 2. Walking with phone in pocket/bag \\ \cline{2-2} 
                                         & 3. Playing games                    \\ \cline{2-2} 
                                         & 4. Browsing                         \\ \cline{2-2} 
                                         
                                         & 5. Making phone calls               \\ \cline{2-2} 
                                         & 6. Making video calls               \\ \bottomrule
\end{tabular}
\caption{Typical Activities of Users on Smart Device~\cite{activity}.}
\label{tasklist}
\end{table}
conditions, the malicious App could open up the camera and record videos surreptitiously. We collected 15 different datasets from these three attack scenarios to test the efficacy of 6thSense against these adversaries. %We collected 15 datasets in total from these three threats.
\begin{table*}[htb!]
\centering
\begin{tabular}{ccccccc}
\toprule
\textbf{\begin{tabular}[c]{@{}c@{}}Threshold\\(Number of consecutive\\ malicious states )\end{tabular}} & \textbf{\begin{tabular}[c]{@{}c@{}}Recall\\   rate\end{tabular}} & \textbf{\begin{tabular}[c]{@{}c@{}}False negative\\ rate\end{tabular}} & \textbf{\begin{tabular}[c]{@{}c@{}}Precision rate \\ (specificity)\end{tabular}}   & \textbf{\begin{tabular}[c]{@{}c@{}}False positive \\rate\end{tabular}} & \textbf{Accuracy} & \textbf{F-score} \\ \hline
\midrule
0                                                                               & 0.62                                                    & 0.38                & 1                           & 0                   & 0.6833   & 0.7654  \\ 
1                                                                               & 0.86                                                    & 0.14                & 1                           & 0                   & 0.8833   & 0.9247  \\ 
\rowcolor{Gray}
2                                                                               & 0.96                                                    & 0.04                & 1                           & 0                   & 0.9667   & 0.9796  \\ 
\rowcolor{Gray}
3                                                                               & 0.98                                                  & 0.02                & 1                           & 0                   & 0.9833   & 0.9899  \\ 
\rowcolor{Gray}
5                                                                               & 1                                                       & 0                   & 0.9                         & 0.1                 & 0.9833   & 0.9474  \\ 
6                                                                               & 1                                                       & 0                   & 0.8                         & 0.2                 & 0.9667   & 0.8889  \\ 
8                                                                               & 1                                                       & 0                   & 0.6                         & 0.4                 & 0.9333   & 0.75    \\ 
10                                                                              & 1                                                       & 0                   & 0.5                         & 0.5                 & 0.9167   & 0.6667  \\ 
12                                                                              & 1                                                       & 0                   & 0.5                         & 0.5                 & 0.9167   & 0.6667  \\ 
15                                                                              & 1                                                       & 0                   & 0.3                         & 0.7                 & 0.8833   & 0.4615  \\ 
\bottomrule
\end{tabular}
\caption{Performance evaluation of Markov Chain based model.}
\label{markovchain}
\end{table*}
% To ensure data quality we stopped all the apps except for the apps that are being accessed by user. 

\subsection{Dataset}
In order to test 6thSense, we divided the collected real user data into two sections as it is a common practice \cite{Radh1310:Passive}. 75\% of the  collected benign dataset was used to train the 6thSense framework and 25\% of the collected data along with malicious dataset were used for testing purposes. For the Markov Chain-based detection technique, the training dataset was used to compute the state transitions and to build the transition matrix. On the other hand, in the Naive Bayes-based detection technique, the training dataset was used to determine the frequency of sensor condition changes for a particular activity or task. As noted earlier, there were nine activities for the Naive Bayes technique. We split the data according to their activity for this approach. For the analysis of the other ML-based approaches, we define all the data in benign and malicious classes. The data were then used to train and test 6thSense using 10-fold cross validation for different ML algorithms.
% I dont get it?
% To ensure randomness, we choose different dataset from different collected user data for user related activities to build the training dataset. 

\subsection{Performance Metrics}
In the evaluation of 6thSense, we utilized the following six different performance metrics: Recall rate (sensitivity or True Positive rate), False Negative rate, Specificity (True Negative rate), False Positive rate, Accuracy, and F-score. True Positive (TP) indicates number of benign activities that are detected correctly while true negative (TN) refers to the number of correctly detected malicious activities. On the other hand, False Positive (FP) states malicious activities that are detected as benign activities and False Negative (FN) defines number of benign activities that are categorized as malicious activity. F-score is the performance metric of a framework that reflects the accuracy of the framework by considering the recall rate and specificity. These performance metrics are defined as follows:

\begin{equation}
{Recall\ rate} = \frac{TP}{TP+FN} ,
\end{equation}
\begin{equation}
{False\ negative\ rate} = \frac{FN}{TP+FN} ,
\end{equation}
\begin{equation}
{Specificity} = \frac{TN}{TN+FP} ,
\end{equation}
\begin{equation}
{False\ positive\ rate} = \frac{FP}{TN+FP} ,
\end{equation}
\begin{equation}
{Recall\ rate} = \frac{TP}{TP+FN} ,
\end{equation}
\begin{equation}
{Accuracy} = \frac{TP+TN}{TP+TN+FP+FN} ,
\end{equation}
\begin{equation}
{F-score} = \frac{2*Recall\ rate*Precision\ rate}{Recall\ rate+Precision\ rate} 
\end{equation}
In addition to the aforementioned performance metrics, we considered Receiver Operating Characterstic (ROC) curve %and Precision Recall Curve (PRC) 
as another performance metric for 6thSense. %As our collected dataset is imbalanced (number of benign events is higher than malicious events), the area under PRC is considered better performance metric to compare between different ML techniques \cite{Roy:2015:ESR:2818000.2818038}. 

%\begin{figure*}[htbp]
%\centering
%\subfloat[False Negative and True Negative Rate]{\includegraphics[height=5cm, width=5.5cm]{false_positive_negative_bayes}}
%\subfloat[Accuracy and F-score]{\includegraphics[height=5cm, width=5.5cm]{Acc_vs_F_naive}}
%\subfloat[ROC Curve]{\includegraphics[height=5cm, width=5.5cm]{ROC_naive}}
%\caption{Performance evaluation of Naive Bayes based approach}
%\label{naivefig}
%\end{figure*}
\begin{table*}[t!]
\centering

\begin{tabular}{ccccccc}
\toprule
\textbf{\begin{tabular}[c]{@{}c@{}}Threshold \\ Probability \end{tabular}}& \textbf{\begin{tabular}[c]{@{}c@{}}Recall\\ rate \end{tabular}} & \textbf{\begin{tabular}[c]{@{}c@{}}False negative\\ rate\end{tabular}} & \textbf{\begin{tabular}[c]{@{}c@{}}Precision rate\\(specificity)\end{tabular}} & \textbf{\begin{tabular}[c]{@{}c@{}}False positive\\ rate\end{tabular}} & \textbf{Accuracy} & \textbf{F-score}   \\ \hline
\midrule
55\%                  & 1                                                       & 0                   & 0.6                         & 0.4                 & 0.9333   & 0.75     \\ 
57\%                  & 1                                                       & 0                   & 0.7                         & 0.3                 & 0.95     & 0.8235   \\ 
\rowcolor{Gray}
60\%                  & 1                                                       & 0                   & 0.7                         & 0.3                 & 0.95     & 0.8235   \\ 
\rowcolor{Gray}
62\%                  & 1                                                       & 0                   & 0.7                         & 0.3                 & 0.95     & 0.8235   \\ 
\rowcolor{Gray}
65\%                  & 0.94                                                    & 0.06                & 0.7                         & 0.3                 & 0.9      & 0.8024   \\ 
67\%                  & 0.88                                                    & 0.12                & 0.7                         & 0.3                 & 0.85     & 0.7797   \\ 
70\%                  & 0.7                                                     & 0.3                 & 0.8                         & 0.2                 & 0.7167   & 0.7467   \\ 
72\%                  & 0.7                                                     & 0.3                 & 0.9                         & 0.1                 & 0.7333   & 0.7875   \\ 
75\%                  & 0.66                                                    & 0.34                & 0.9                         & 0.1                 & 0.7      & 0.7616 \\ 
80\%                  & 0.66                                                    & 0.34                & 0.9                         & 0.1                 & 0.7      & 0.7615   \\ 
\bottomrule
\end{tabular}
\caption{Performance evaluation of Naive Bayes model.}
\label{naivetable}
\end{table*}
\subsection{Evaluation of Markov Chain-Based Detection}

In the Markov Chain-based detection technique, we question whether the transition between two states (sensors' on/off condition in each second) is expected or not. In the evaluations, we used 65 testing sessions in total, among which 50 sessions were for the  benign activities and the rest of the sessions were for the malicious activities. A session is composed of a series of sensory context conditions where a sensory context condition is the set of all available sensor conditions (on/off state) for different sensors. As discussed earlier in Section 6, a sensor condition is a value indicating whether the sensor data is changing or not. 
In this evaluation, the sensory context conditions were computed every one second. We observed that in real devices sometimes some sensor readings would be missed or real data would not be reflected probably due to hardware or software imperfections. %Such  faulty sensor readings would cause malicious states in the system for limited time instances. %but it will not be consecutive. Otherwise, sensor implemented on the device will be considered as malfunctioning for giving faulty data. 
%
%On the other hand, 
And, real malicious Apps would cause consecutive malicious states on the device. Therefore, to overcome this, we also keep track of number of consecutive malicious states and use it as a threshold after which the session is considered as malicious.
Table~\ref{markovchain} displays the evaluation results associated wit the Markov Chain-based detection technique.  When the threshold for consecutive malicious states is 0, i.e., when no threshold is  applied, the accuracy is just 68\% and FNR is as high as 38\%. With increasing the threshold value, the accuracy first increases up to 98\% then start decreasing. %An optimal results are obtained for the values of 

\begin{figure}[h!]
  \centering
  \framebox{
    \includegraphics[height=5cm, width=6cm]{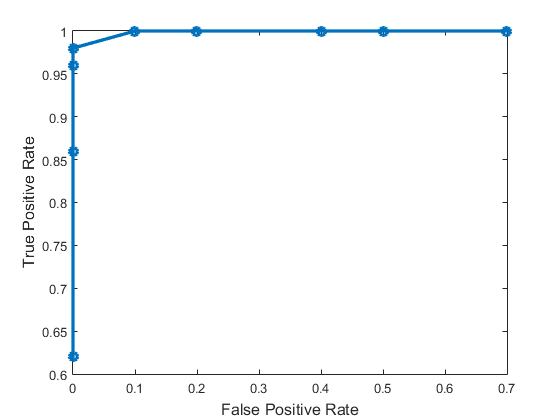}
    }
      \caption{ROC curve of Markov Chain-based detection.}
      \label{markovfig}
\end{figure}

% From table 2, we can see that considering only multiplication rule give a higher false negative rate which is 0.38. Though there is no false positive case for this method, due to higher false negative rate, accuracy becomes much smaller. To mitigate this, we consider the number consecutive malicious state as a threshold in our framework. We define this parameter as following --- if there are consecutive occurrences of malicious state, we will consider it as malicious behavior. As we observe state change in sensor in a per second manner, we can assume that, a malicious attack will need more than one second to execute a task in the device. From table 2, we observe that, if we consider one consecutive occurrence of malicious state, false negative rate improves from 0.38 to 0.14. If we increase the threshold of consecutive malicious occurrence, false negative rate decreases. But as a trade off, it introduces false positive rate in the framework. If we consider 5 consecutive malicious state occurrences as valid occurrence in our framework, accuracy is the highest. At the same time, it introduces false positive rate in the framework which is much severe from the security perspective. 
The possible cut-off threshold can be three consecutive malicious occurrences which has both accuracy and F-score over 98\%. In Table~\ref{markovchain}, different performance indicators for Markov Chain based detection are presented. We can observe that FN and TN rates of Markov Chain-based detection decrease as the threshold of consecutive malicious states increases.  Again, both accuracy and F-score reach to a peak value with the threshold of three consecutive malicious states on the device. From Figure~\ref{markovfig}, we can see that FP rate remains zero while TP rate increases at the beginning. The highest TP rate without introducing any FP  case is over 98\%. After 98\%, it introduces some FP cases in the system which is considered as a risk to the system. In summary, Markov Chain-base detection in 6thSense can acquire accuracy over 98\% without introducing any FP cases.

% As we consider transition between two states as independent events, we use multiplication rule to calculate total probability which indicates the validation of a session. 

% we might chose one of words: approach, method or technique and stick to it in all paper.

\subsection{Evaluation of Naive Bayes-based Detection}
In the Naive Bayes-based detection technique, 6thSense calculates the probability of a session to match it with each activity defined in Section\ref{sec:trainingenv}. Since all the activities are benign and there is no malicious activity (i.e., ground-truth data), 6thSense checks calculated probability of an activity from dataset against a threshold to determine the correct activity. If there is no match for a certain sensor condition with any of the activities, 6thSense detects the session as malicious. Table~\ref{naivetable} shows the evaluation results. 

For a threshold value of 55\%, FN rate is zero. However, FPR is too high, which lowers F-score of the framework. For a threshold of 60\%, FPR decreases while FNR is still zero. In this case, accuracy is 95\% and F-score is 82\%. If the threshold is increased over 65\%, it reduces the recall rate which affects accuracy and F-score. The evaluation indicates that the threshold value of 60\% provides an accuracy of 95\% and F-score of 82\%.

\begin{figure}[h!]
  \centering
  \framebox{
    \includegraphics[height=5cm, width=6cm]{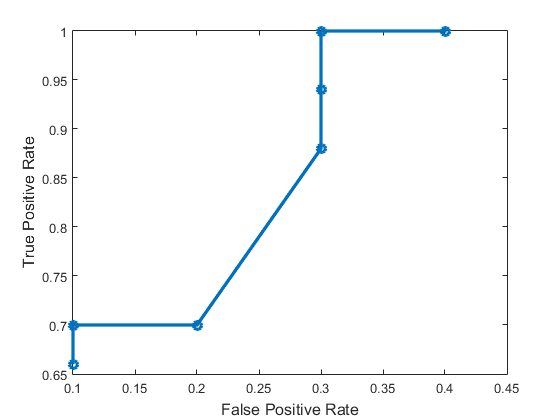}
   }
      \caption{ROC curve of Naive Bayes-based detection.}
      \label{naivefig}
\end{figure}

%Figure~\ref{naivefig} displays different performance metrics for Naive Bayes based approach. Figure~\ref{naivefig}(a) shows that both false negative and true negative increases with increment of threshold value. After threshold of 0.75, both false negative and true negative rate become constant. Figure~\ref{naivefig}(b) indicates that both accuracy and F-score remain highest between 0.57 to 0.62 threshold then gradually decreases. 
From Figure~\ref{naivefig}, one  can observe the relation between FPR and TPR of Naive Bayes-based detection. For FPR larger than 0.3, TPR becomes 1.

\subsection{Evaluation of Alternative Detection Techniques}
In alternative detection techniques, we used other supervised machine learning techniques to train the 6thSense  framework. For this, we utilized WEKA and it provides three types of analysis - split percentage analysis, cross-validation analysis, and supplied test set analysis. We chose 10 fold cross-validation analysis to ensure that all the data was used for both training and test. Thus, the error rate of the predictive model would be minimized in the cross validation. In Table~\ref{wekatable}, a detailed evaluation of different machine learning algorithms is given for 6thSense. For \textit{Rule Based Learning}, 6thSense has the best result for \textit{PART} algorithm, which has an accuracy of 0.99 and F-score of 0.7899. On the other hand, for \textit{Regression Analysis}, we use the logistic function which has high FPR (0.7222) and lower F-score (0.4348). 
\textit{Multilayer Perceptron} algorithm gives an accuracy of 0.9991 and F-score of 0.8196,  which is higher than previously mentioned algorithms. However, FPR is much higher (0.3056), which is actually a limitation for intrusion detection frameworks in general. Compared to these algorithms, \textit{Linear Model Tree (LMT)} gives better results in detecting sensor-based attacks. This evaluation indicates that \textit{LMT} provides an accuracy of 0.9997 and F-score of 0.964.  
\begin{table*}[t!]
\centering
\begin{tabular}{ccccccc}
\toprule
\textbf{Algorithms} & \textbf{\begin{tabular}[c]{@{}l@{}}Recall\\rate\end{tabular}} & \textbf{\begin{tabular}[c]{@{}l@{}}False negative\\ rate\end{tabular}} & \textbf{Precision rate} & \textbf{\begin{tabular}[c]{@{}l@{}}False positive\\ rate\end{tabular}} & \textbf{Accuracy} & \textbf{F-score}  \\ \hline
\midrule
PART         & 0.9998          & 0.0002         & 0.6528     & 0.3472     & 0.99   & 0.7899     \\ 
Logistic Function    & 0.9997  & 0.0003         & 0.2778     & 0.7222   & 0.998    & 0.4348   \\ 
J48         & 0.9998          & 0.0002         & 0.6528     & 0.3472     & 0.99   & 0.7899   \\ 
\rowcolor{Gray}
LMT           & 0.9998          & 0.0002       & 0.9306            & 0.0694        & 0.9997     & 0.964   \\ 
Hoeffding Tree   & 1     & 0     & 0.0556    & 0.9444        & 0.9978     & 0.1053   \\ 
Multilayer Perceptron     & 0.9998      & 0.0002    & 0.6944      & 0.3056   & 0.9991  & 0.8196   \\ \bottomrule

\end{tabular}
\caption{Performance of other different machine learning based-detection techniques tested in 6thSense.}
\label{wekatable}
\end{table*}

\subsection{Comparison}

In this subsection, we give a comparison among the different machine-learning-based detection % the three 
approaches tested in 6thSense %we proposed 
for defending against sensor-based threats. For all the approaches, we select the best possible case and report their performance metrics in Table~\ref{comparison}. For Markov Chain-based detection, we choose three consecutive malicious states as valid device conditions. On the other hand, in Naive Bayes approach, the best performance is observed for the threshold of 60\%. For other machine learning algorithms tested via WEKA, we choose LMT as it gives highest accuracy among other machine learning algorithms. These results indicate that LMT provides highest accuracy and F-score compared to the other two approaches.

\par On the contrary, Naive Bayes model displays higher recall rate and less FNR than other approaches. However, the presence of FPR in IDS is a serious security threat to the system since FPR refers to a malicious attack that is identified as a valid state, which is a threat to user privacy and security of the device. Both Markov Chain and LMT has lower FPR. %\action{However, if we consider auPRC of three approaches, Markov Chain performs better, followed by LMT. In summary, considering F-score, accuracy, and auPRC of all three approaches, we conclude that Markov Chain and LMT both performs effectively for 6thSense.}
In summary, considering F-score and accuracy of all these approaches, we conclude that LMT performs better than the others. 
\begin{table}[h!]
\centering
\fontsize{8}{10}\selectfont
\begin{tabular}{lllll}
\toprule
\textbf{\begin{tabular}[c]{@{}l@{}}Performance \\Metrics\end{tabular}} & \textbf{\begin{tabular}[c]{@{}l@{}}Markov \\Chain\\  \end{tabular}} & \textbf{\begin{tabular}[c]{@{}l@{}}Naive \\Bayes\end{tabular}}  & \textbf{LMT} \\ \hline
\midrule
Recall rate         & 0.98      & 1         & 0.9998                 \\ 
\begin{tabular}[c]{@{}l@{}}False Negative Rate \end{tabular}& 0.02      & 0         & 0.0002                \\
Precision rate      & 1         & 0.7       & 0.9306                \\ 
\begin{tabular}[c]{@{}l@{}}False positive rate \end{tabular}& 0         & 0.3       & 0.0694                \\
Accuracy            & 0.9833    & 0.9492    & 0.9997                \\
F-Score             & 0.9899    & 0.8235    & 0.964                \\
auPRC               & 0.947     & 0.686     & 0.91                  \\
\bottomrule
\end{tabular}
\caption{Comparison of different machine-learning-based approaches proposed for 6thSense (i.e., Markov Chain, Naive Bayes, and LMT).}
\label{comparison}
\end{table}

%\begin{figure}[h!]
%  \centering
%  \framebox{
 % 
    % \includegraphics[width=\textwidth,height=8cm]{new3-1.png}
  %  \includegraphics[width=7.5cm,height=6cm]{Comparison}
   % }
    %  \caption{Performance comparison between Markov Chain based and Naive Bayes based approach}
% \label{comparisonfig}
%\end{figure}
% Figure~\ref{comparisonfig} gives a graphical comparison of all the performance metrics of Markov Chain and naive Bayes based approach.  
\subsection{Performance Overhead}
As previously mentioned, 6thSense collects data in an Android device from different sensors (permission and no-permission imposed sensors). In this sub-section, we measure the performance overhead introduced by 6thSense on the tested Android device in terms of CPU usage, RAM usage, file size, and power consumption and Table~\ref{overhead} gives the details of the performance overhead.

For no-permission-imposed sensors, the data collection phase %collector 
logs all the values within a time interval which causes an increased usage of RAM, CPU and Disc compared to permission- imposed or logic-oriented sensors. For the power consumption, we observe that  no-permission-imposed sensors use higher power than permission-imposed sensors. This is mainly because logic-oriented sensors have lower sampling rate, which reduces its resource needs.

The overall performance overhead is as low as 4\% of CPU, less then $40MB$ RAM space, and less than $15MB$ disc space. Thus, its overhead is minimal and acceptable for an IDS system on current smartphones. 
% On the other hand, sensor logic state detector uses much less CPU and RAM space than the other one. Because of lower sampling rate of logic oriented sensors, disc usage is also small and power consumption is much less. Table 5 gives detail overview of performance overheads of our customized android apps.
One of the main concerns of implementing 6thSense on Android device is the power consumption.

Table~\ref{overhead} also shows the power consumption of the Android app used in 6thsense. For one minute, 6thsense consumes 16.62 mW power which increases upto 178.33mW for ten seconds. The main reason of this high power consumption is that  all the sensors are kept on for the data collection and all the data are saved on device for later analysis. However, in practical settings, the data would not be saved on device rather a real time analysis would  be done, which indeed will decrease the power consumption. Without saving the data, the power consumption significantly becomes smaller. From Table~\ref{overhead}, we can observe that the power consumption of 6thSense becomes 72.35 mW which is almost 2.5 times lower than otherwise. Also, all the sensors do not have to remain on for the analysis part. Data can be observed if the smart device is in unlocked status. Also, a suitable interval can be chosen for the data analysis by estimating average time of an attack. This is one of the possible future research directions for 6thSense. 
\begin{table}[h!]
\centering
\fontsize{8}{10}\selectfont
\begin{tabular}{p{2cm}p{1.2cm}p{1.6cm}p{1.5cm}}
\toprule
\textbf{Parameters}                                                                    & \textbf{Time}       & \textbf{\begin{tabular}[c]{@{}l@{}}No-permission\\imposed \\sensors\end{tabular}} & \textbf{\begin{tabular}[c]{@{}l@{}}Permission\\imposed\\ sensors\end{tabular}} \\ \hline
\midrule
CPU Usage                                                                     & N/A        & 3.90\%                                                           & 0.3\%                                                                 \\ \hline
RAM Usage                                                                     & N/A        & 23 MB                                                            & 14 MB                                                                 \\ \hline
\multirow{3}{*}{Disc Usage}                                                    & For 1 min  & 6.5 MB                                                           & 1 KB                                                                  \\ \cline{2-4} 
                                                                              & For 5 min  & 9 MB                                                             & 2 KB                                                                  \\ \cline{2-4} 
                                                                              & For 10 min & 12 MB                                                            & 3 KB                                                                  \\ \hline
\multirow{3}{*}{\begin{tabular}[p{1.7cm}]{@{}p{2cm}@{}}Power\\Consumption\end{tabular}} & 1 min      & 13.5 mW                                                           & 3.12 mW                                                                \\ \cline{2-4} 
                                                                              & 5 min      & 96.67 mW                                                           & 27.4 mW                                                                 \\ \cline{2-4} 
                                                                              & 10 min     & 133.33 mW                                                             & 45 mW                                                                 \\ \hline
\multirow{3}{*}{\begin{tabular}[p{1.7cm}]{@{}p{2cm}@{}}Power\\Consumption\\(without datafile)\end{tabular}} & 1 min      & 2.68 mW                                                           & 0.23 mW                                                                \\ \cline{2-4} 
                                                                              & 5 min      &  23.4 mW                                                           & 9.63 mW                                                                 \\ \cline{2-4} 
                                                                              & 10 min     & 55.35 mW                                                             & 17 mW                                                                 \\ \hline
\bottomrule                                                                              
\end{tabular}
\caption{Performance Overhead of Android Apps.}
\label{overhead}
\end{table}
\section{Discussion and Limitations}\label{sec:Discussion}
\begin{itemize}
    
\item \textbf{Features and Benefits-} Compared to the existing solutions,  %to secure sensors on smartphone, 
6thSense differentiates itself by considering a context-aware model to detect sensor-based threats. As sensors provide continuous data to the apps, security schemes must handle real-time data rather than stored data in the system. While most of the existing solutions work with the stored data or the data used by Apps~\cite{bugiel2011practical,hilty2007policy}, 6thSense offers real-time sensor monitoring. On the other hand, modern high precision sensors on-board have higher frequency and sensitivity. These sensors can detect slight changes in the smart device's ambiance which reflects on sensor values. To overcome frequent changes in sensor values, 6thSense considers average values over one second, which mitigates the effect of changes in sensor values caused by the device ambiance. For example, if a person walks by a smartphone, the light sensor and motion sensors values will be changed for that instance. However, if one  considers the average value over one second, it will be compensated by other readings recorded over one second. Another unique feature of 6thSense is that instead of considering the  individual sensor data accessed by the Apps, user activities are monitored, which forms the basis of the contextual model for the 6thSense  framework. 6thSense observes changes in sensors for different user activities. As more than one sensor remain active to perform a task, attackers need to learn the pattern of all the sensors for user activities to outperform 6thSense. If an attacker targets one specific sensor, an attack scenario will differ from normal user activity which can be easily detected by 6thSense. Thus, the context of user activities is very important to detect malicious activities in 6thSense.  Moreover, 6thSense considers all the sensors' conditions as one device state, which provides easy monitoring of the sensors by one framework. %{\color{red}6thSense also offers different detection techniques in the framework. Our detailed evaluation of different detection techniques aids to identify the optimal analytical model for 6thSense. This knowledge can help researchers to learn about the effectiveness of different detection techniques to mitigate sensor-based threats.} 
Finally, 6thSense can work with all the sensors 
%Thus, our framework covers up all the sensors 
on a smart device extending the security beyond the traditional permission-imposed sensors (i.e., GPS, microphone, and camera). 
%\textbf{Comparison with Prior Works.} There are different solutions proposed by researchers addressing problems on sensor management systems. 
%\textbf{Power Consumption.} 
\item \textbf{Application Level Detection-} One of the promising practical applications %future research areas for 
of 6thSense is to combine the sensor level detection with an application level intrusion detection. 6thSense focuses on detecting malicious activities by observing working conditions of sensors rather than individual App behaviors. However, some prior works \cite{Xu:2015:SPS:2699026.2699114, Petracca:2015:APA:2818000.2818005, Wang:2015:NHM:2802130.2802132} also show that it is possible to achieve good accuracy when  detecting malicious activities by observing sensor usage in the application level. The combination of application and sensor level detection might be one promising way to further improve the performance of 6thSense. Another interesting application of 6thSense would be to combine it with an online training method to eliminate the necessity of offline training.
%Different online training methods can be explored to eliminate the necessity of offline training for 6thSense to offer smooth operation

%and real-life malware

\item \textbf{Sensor-based threats in real-life settings -} One limitation of 6thSense is the adversaries (sensor-based attacks) used in the evaluation were  constructed in a lab environment. Note that as of this writing, there are no real sensor-based malware in the wild.  
%The reason of using lab-made adversary model is the lack of sensor-based malicious apps that attacks smart devices via sensor in real world. 
However, recently, many independent researchers have confirmed the feasibility of sensor-based threats for smart devices \cite{Petracca:2015:APA:2818000.2818005, carlini2016hidden, enigma, son2015rocking}. Indeed, more recently, ICS-CERT also warned the vendors and the wider communities about the possibility of  
%asserted that it is possible to 
exploiting the sensors of a device to alter sensors’ output in a controlled way to perform malicious behavior in the device \cite{ICS-CERT}. %Again, there are malware that can achieve the same goal as sensor-based threats by exploiting different channels.
Although there are different limited security schemes to mitigate these attacks, there is no comprehensive contextual solution to secure smart devices from the sensor-based threats. Furthermore, we note that even locking down the sensor API with explicit permissions at the OS level would not surpass the  sensor-based threats as users are not aware of these threats yet and can allow malicious Apps to use sensors unknowingly. For all these points, we built the prof-of-concept  versions of the sensor-based threats discussed in Section 4. %These increasing threats and the lack of real-life sensor-based malicious Apps encouraged us to build a prof-of-concept synthetic/lab-made adversary models. 
We also note that to ensure the reliability of the lab-made malware (i.e., a specific malicious App) for three threats described in the Adversary Model Section,  we checked how they perform with respect to the real malicious software scanners. For this, we uploaded our lab-made malware on \textit{VirusTotal} and tabulated the results of the performance of 60 different malware scanners available at the VirusTotal website in Table~\ref{scan}. As seen in this table, the sensor-based threats are not recognized by the different scanners. Only 2 out of 60 reported that they could detect, but these two only reported risks without clearly identifying any explicit malicious behaviour. 
\begin{table}[h!]
\centering
\fontsize{8}{10}\selectfont
\begin{tabular}{cc}
\toprule
\textbf{Adversary Model} & \textbf{Detection Ratio} \\ \hline
\midrule
Threat-1 & 2/60 \\\hline
Threat-2 & 2/60 \\\hline
Threat-3 & 3/62 \\\hline
\bottomrule                                     
\end{tabular}
\caption{\textit{VirusTotal} scan result for the adversary models.}
\label{scan}
\end{table}
Hence, it is difficult to detect the sensor-based threats mentioned in this paper by existing security schemes. Moreover, some security schemes only provide security to the specific sensors \cite{Petracca:2015:APA:2818000.2818005}. 6thSense covers several sensors as opposed to other existing existing security schemes without alerting the device. Also, existing sensor management systems of Android depends on explicit user-permission only for specific sensors (e.g., microphone, camera, speaker). As users are not aware of sensor-based threats yet, they can allow malicious Apps to use sensors unknowingly. Additionally, 6thSense also covers no-permission-imposed sensors (e.g., motion sensors, light sensor, etc.) in its design.  %Hence, it is difficult for existing schemes to detect our adversary models efficiently.  

\item \textbf{Context-aware Malicious App-} One compelling case is that how 6thSense can defend against a malicious App which can learn and imitate a user’s behavior. %In our adversary model, we address this question by creating a malicious app that can observe sensors’ working conditions. 
As described earlier, Threat 3, described in Section 4.1, can observe the working conditions of all the sensors and detect, for instance, a sleeping activity that records videos stealthily. 6thSense can even detect this powerful context-aware malware successfully. In summary, to outperform 6thSense, a malicious App must behave like a benign App all the time in a device,  which limits its malicious purposes. Any incompatible behavior in the sensors of a smart device can be easily detected by 6thSense.
\end{itemize}
\ROneEnd

\section{Conclusion}\label{sec:Conclusion}
Wide utilization of sensor-rich smart devices created a new attack surface namely sensor-based attacks. Accelerometer, gyroscope, light, etc. sensors can be abused to steal and leak sensitive information or malicious Apps can be triggered via sensors. Security in current smart devices lacks appropriate defense mechanisms for such sensor-based threats. In this paper, we presented 6thSense, a novel context-aware task-oriented sensor-based attack detector for smart devices. We articulated problems in existing sensor management systems and different sensor-based threats for smart devices. Then, we presented the design of 6thSense to detect sensor-based attacks on a sensor-rich smart device with low-performance overhead. 6thSense utilized different machine learning (ML) techniques to distinguish malicious activities from benign activities on a device. To the best of our knowledge, 6thSense is the first comprehensive context-aware security solution against sensor-based threats. We evaluated 6thSense on real devices with 50 different individuals. 6thSense achieved over 95\% of accuracy with different ML algorithms including Markov Chain, Naive Bayes, and LMT . We also evaluated 6thSense against three different sensor-based threats, i.e., information leakage, eavesdropping, and triggering a malware via sensors. The empirical evaluation revealed that 6thSense is highly effective and efficient at detecting sensor-based attacks while yielding minimal overhead. 

\textit{\textbf{Future Work:}}  While 6thSense detects different sensor-based threats with high accuracy, we will expand 6thSense in our future work as follows: We will study other performance metrics such as Precision Recall Curve (PRC). We will evaluate  the efficacy of 6thSense in other smart devices such as smartwatches and analyze all of its phases in its operations. Moreover, due to limited resources of the smart devices, trade-off between power consumption and effectiveness is a prime concern of any intrusion detection framework. Hence, we will study % Further research is needed to 
frequency-accuracy trade-off, battery-accuracy trade-off, and battery-frequency trade-off of  6thSense in different smart devices. %We also will release the collected data from users for this work.

{\footnotesize \bibliographystyle{acm}
\bibliography{sample}}

\begin{thebibliography}{10}

\bibitem{enigma}
Hacking sensors.
\newblock
  \url{https://www.usenix.org/conference/enigma2017/conference-program/presentation/kim}.
\newblock Accessed: 2017-5-30.

\bibitem{ICS-CERT}
Mems accelerometer hardware design flaws (update a).
\newblock \url{https://ics-cert.us-cert.gov/alerts/ICS-ALERT-17-073-01A}.
\newblock Accessed: 2017-5-30.

\bibitem{activity}
U.s. smartphone use in 2015.
\newblock
  \url{http://www.pewinternet.org/2015/04/01/us-smartphone-use-in-2015/}, April
  2015.

\bibitem{activity4}
A week in the life analysis of smartphone users.
\newblock \url{http://www.pewinternet.org/2015/04/01/}, April 2015.

\bibitem{antivirus}
Analyzing the power consumption of mobile antivirus software on android
  devices.
\newblock \url{http://drshem.com/2015/11/08/}, August 2016.

\bibitem{norton}
Android antivirus protection: Security steps you should take.
\newblock \url{http://us.norton.com/Android-Anti-Virus-Protection/article}, sep
  2016.

\bibitem{androidmarket}
Smartphone os market share, 2016 q2.
\newblock \url{http://www.idc.com/prodserv/smartphone-os-market-share.jsp},
  August 2016.

\bibitem{Samsung}
Smartphone vendor market share, 2016 q2.
\newblock \url{http://www.idc.com/prodserv/smartphone-market-share.jsp}, August
  2016.

\bibitem{al2013keystrokes}
{\sc Al-Haiqi, A., Ismail, M., and Nordin, R.}
\newblock Keystrokes inference attack on android: A comparative evaluation of
  sensors and their fusion.
\newblock {\em Journal of ICT Research and Applications 7}, 2 (2013), 117--136.

\bibitem{1301311}
{\sc Asonov, D., and Agrawal, R.}
\newblock Keyboard acoustic emanations.
\newblock In {\em Security and Privacy, 2004. Proceedings. 2004 IEEE Symposium
  on\/} (May 2004), pp.~3--11.

\bibitem{aviles2011comparison}
{\sc Avil{\'e}s-Arriaga, H., Sucar-Succar, L., Mendoza-Dur{\'a}n, C., and
  Pineda-Cort{\'e}s, L.}
\newblock A comparison of dynamic naive bayesian classifiers and hidden markov
  models for gesture recognition.
\newblock {\em Journal of applied research and technology 9}, 1 (2011),
  81--102.

\bibitem{Aviv:2012:PAS:2420950.2420957}
{\sc Aviv, A.~J., Sapp, B., Blaze, M., and Smith, J.~M.}
\newblock Practicality of accelerometer side channels on smartphones.
\newblock In {\em Proceedings of the 28th Annual Computer Security Applications
  Conference\/} (New York, NY, USA, 2012), ACSAC '12, ACM, pp.~41--50.

\bibitem{brooks2011handbook}
{\sc Brooks, S., Gelman, A., Jones, G., and Meng, X.-L.}
\newblock {\em Handbook of Markov Chain Monte Carlo}.
\newblock CRC press, 2011.

\bibitem{bugiel2011practical}
{\sc Bugiel, S., Davi, L., Dmitrienko, A., Heuser, S., Sadeghi, A.-R., and
  Shastry, B.}
\newblock Practical and lightweight domain isolation on android.
\newblock In {\em Proceedings of the 1st ACM workshop on Security and privacy
  in smartphones and mobile devices\/} (2011), ACM, pp.~51--62.

\bibitem{Cai:2011:TIK:2028040.2028049}
{\sc Cai, L., and Chen, H.}
\newblock Touchlogger: Inferring keystrokes on touch screen from smartphone
  motion.
\newblock In {\em Proceedings of the 6th USENIX Conference on Hot Topics in
  Security\/} (Berkeley, CA, USA, 2011), HotSec'11, USENIX Association,
  pp.~9--9.

\bibitem{cai2012practicality}
{\sc Cai, L., and Chen, H.}
\newblock {\em On the practicality of motion based keystroke inference attack}.
\newblock Springer, 2012.

\bibitem{carlini2016hidden}
{\sc Carlini, N., Mishra, P., Vaidya, T., Zhang, Y., Sherr, M., Shields, C.,
  Wagner, D., and Zhou, W.}
\newblock Hidden voice commands.
\newblock In {\em 25th USENIX Security Symposium (USENIX Security 16), Austin,
  TX\/} (2016).

\bibitem{chan2009smart}
{\sc Chan, M., Campo, E., Est{\`e}ve, D., and Fourniols, J.-Y.}
\newblock Smart homes—current features and future perspectives.
\newblock {\em Maturitas 64}, 2 (2009), 90--97.

\bibitem{coffed2014threat}
{\sc Coffed, J.}
\newblock The threat of gps jamming: The risk to an information utility.
\newblock {\em Report of EXELIS, Jan. Chicago\/} (2014).

\bibitem{dahl2013large}
{\sc Dahl, G.~E., Stokes, J.~W., Deng, L., and Yu, D.}
\newblock Large-scale malware classification using random projections and
  neural networks.
\newblock In {\em Acoustics, Speech and Signal Processing (ICASSP), 2013 IEEE
  International Conference on\/} (2013), IEEE, pp.~3422--3426.

\bibitem{Diao:2014:YVA:2666620.2666623}
{\sc Diao, W., Liu, X., Zhou, Z., and Zhang, K.}
\newblock Your voice assistant is mine: How to abuse speakers to steal
  information and control your phone.
\newblock In {\em Proceedings of the 4th ACM Workshop on Security and Privacy
  in Smartphones \&\#38; Mobile Devices\/} (New York, NY, USA, 2014), SPSM '14,
  ACM, pp.~63--74.

\bibitem{Enck:2014:TIT:2642648.2619091}
{\sc Enck, W., Gilbert, P., Han, S., Tendulkar, V., Chun, B.-G., Cox, L.~P.,
  Jung, J., McDaniel, P., and Sheth, A.~N.}
\newblock Taintdroid: An information-flow tracking system for realtime privacy
  monitoring on smartphones.
\newblock {\em ACM Trans. Comput. Syst. 32}, 2 (June 2014), 5:1--5:29.

\bibitem{farooqi2013novel}
{\sc Farooqi, A.~H., Khan, F.~A., Wang, J., and Lee, S.}
\newblock A novel intrusion detection framework for wireless sensor networks.
\newblock {\em Personal and ubiquitous computing 17}, 5 (2013), 907--919.

\bibitem{FooKune:2010:TAP:1866307.1866395}
{\sc Foo~Kune, D., and Kim, Y.}
\newblock Timing attacks on pin input devices.
\newblock In {\em Proceedings of the 17th ACM Conference on Computer and
  Communications Security\/} (New York, NY, USA, 2010), CCS '10, ACM,
  pp.~678--680.

\bibitem{gu2007bothunter}
{\sc Gu, G., Porras, P.~A., Yegneswaran, V., Fong, M.~W., and Lee, W.}
\newblock Bothunter: Detecting malware infection through ids-driven dialog
  correlation.
\newblock In {\em Usenix Security\/} (2007), vol.~7, pp.~1--16.

\bibitem{Halevi:2012:CLK:2414456.2414509}
{\sc Halevi, T., and Saxena, N.}
\newblock A closer look at keyboard acoustic emanations: Random passwords,
  typing styles and decoding techniques.
\newblock In {\em Proceedings of the 7th ACM Symposium on Information, Computer
  and Communications Security\/} (New York, NY, USA, 2012), ASIACCS '12, ACM,
  pp.~89--90.

\bibitem{hall2009weka}
{\sc Hall, M., Frank, E., Holmes, G., Pfahringer, B., Reutemann, P., and
  Witten, I.~H.}
\newblock The weka data mining software: an update.
\newblock {\em ACM SIGKDD explorations newsletter 11}, 1 (2009), 10--18.

\bibitem{Hasan:2013:SCH:2484313.2484373}
{\sc Hasan, R., Saxena, N., Haleviz, T., Zawoad, S., and Rinehart, D.}
\newblock Sensing-enabled channels for hard-to-detect command and control of
  mobile devices.
\newblock In {\em Proceedings of the 8th ACM SIGSAC Symposium on Information,
  Computer and Communications Security\/} (New York, NY, USA, 2013), ASIA CCS
  '13, ACM, pp.~469--480.

\bibitem{hilty2007policy}
{\sc Hilty, M., Pretschner, A., Basin, D., Schaefer, C., and Walter, T.}
\newblock A policy language for distributed usage control.
\newblock In {\em European Symposium on Research in Computer Security\/}
  (2007), Springer, pp.~531--546.

\bibitem{ioannis2007towards}
{\sc Ioannis, K., Dimitriou, T., and Freiling, F.~C.}
\newblock Towards intrusion detection in wireless sensor networks.
\newblock In {\em Proc. of the 13th European Wireless Conference\/} (2007),
  pp.~1--10.

\bibitem{jana2013scanner}
{\sc Jana, S., Narayanan, A., and Shmatikov, V.}
\newblock A scanner darkly: Protecting user privacy from perceptual
  applications.
\newblock In {\em Security and Privacy (SP), 2013 IEEE Symposium on\/} (2013),
  IEEE, pp.~349--363.

\bibitem{jang2014a11y}
{\sc Jang, Y., Song, C., Chung, S.~P., Wang, T., and Lee, W.}
\newblock A11y attacks: Exploiting accessibility in operating systems.
\newblock In {\em Proceedings of the 2014 ACM SIGSAC Conference on Computer and
  Communications Security\/} (2014), ACM, pp.~103--115.

\bibitem{jha2001markov}
{\sc Jha, S., Tan, K.~M., and Maxion, R.~A.}
\newblock Markov chains, classifiers, and intrusion detection.
\newblock In {\em csfw\/} (2001), vol.~1, Citeseer, p.~206.

\bibitem{joy2011side}
{\sc Joy~Persial, G., Prabhu, M., and Shanmugalakshmi, R.}
\newblock Side channel attack-survey.
\newblock {\em Int J Adva Sci Res Rev 1}, 4 (2011), 54--57.

\bibitem{keilson2012markov}
{\sc Keilson, J.}
\newblock {\em Markov chain models—rarity and exponentiality}, vol.~28.
\newblock Springer Science \& Business Media, 2012.

\bibitem{kruegel2003bayesian}
{\sc Kruegel, C., Mutz, D., Robertson, W., and Valeur, F.}
\newblock Bayesian event classification for intrusion detection.
\newblock In {\em Computer Security Applications Conference, 2003. Proceedings.
  19th Annual\/} (2003), IEEE, pp.~14--23.

\bibitem{5560598}
{\sc Lane, N.~D., Miluzzo, E., Lu, H., Peebles, D., Choudhury, T., and
  Campbell, A.~T.}
\newblock A survey of mobile phone sensing.
\newblock {\em IEEE Communications Magazine 48}, 9 (Sept 2010), 140--150.

\bibitem{lane2011enabling}
{\sc Lane, N.~D., Xu, Y., Lu, H., Hu, S., Choudhury, T., Campbell, A.~T., and
  Zhao, F.}
\newblock Enabling large-scale human activity inference on smartphones using
  community similarity networks (csn).
\newblock In {\em Proceedings of the 13th international conference on
  Ubiquitous computing\/} (2011), ACM, pp.~355--364.

\bibitem{6680832}
{\sc Lei, L., Wang, Y., Zhou, J., Zha, D., and Zhang, Z.}
\newblock A threat to mobile cyber-physical systems: Sensor-based privacy theft
  attacks on android smartphones.
\newblock In {\em Trust, Security and Privacy in Computing and Communications
  (TrustCom), 2013 12th IEEE International Conference on\/} (July 2013),
  pp.~126--133.

\bibitem{linda2009neural}
{\sc Linda, O., Vollmer, T., and Manic, M.}
\newblock Neural network based intrusion detection system for critical
  infrastructures.
\newblock In {\em Neural Networks, 2009. IJCNN 2009. International Joint
  Conference on\/} (2009), IEEE, pp.~1827--1834.

\bibitem{s131217292}
{\sc Macias, E., Suarez, A., and Lloret, J.}
\newblock Mobile sensing systems.
\newblock {\em Sensors 13}, 12 (2013), 17292.

\bibitem{maiti2015smart}
{\sc Maiti, A., Jadliwala, M., He, J., and Bilogrevic, I.}
\newblock (smart) watch your taps: side-channel keystroke inference attacks
  using smartwatches.
\newblock In {\em Proceedings of the 2015 ACM International Symposium on
  Wearable Computers\/} (2015), ACM, pp.~27--30.

\bibitem{Meng:2015:CMI:2732198.2732205}
{\sc Meng, W., Lee, W.~H., Murali, S., and Krishnan, S.}
\newblock Charging me and i know your secrets!: Towards juice filming attacks
  on smartphones.
\newblock In {\em Proceedings of the 1st ACM Workshop on Cyber-Physical System
  Security\/} (New York, NY, USA, 2015), CPSS '15, ACM, pp.~89--98.

\bibitem{184479}
{\sc Michalevsky, Y., Boneh, D., and Nakibly, G.}
\newblock Gyrophone: Recognizing speech from gyroscope signals.
\newblock In {\em 23rd USENIX Security Symposium (USENIX Security 14)\/} (San
  Diego, CA, Aug. 2014), USENIX Association, pp.~1053--1067.

\bibitem{milette2012professional}
{\sc Milette, G., and Stroud, A.}
\newblock {\em Professional Android sensor programming}.
\newblock John Wiley \& Sons, 2012.

\bibitem{Miluzzo:2012:TYF:2307636.2307666}
{\sc Miluzzo, E., Varshavsky, A., Balakrishnan, S., and Choudhury, R.~R.}
\newblock Tapprints: Your finger taps have fingerprints.
\newblock In {\em Proceedings of the 10th International Conference on Mobile
  Systems, Applications, and Services\/} (New York, NY, USA, 2012), MobiSys
  '12, ACM, pp.~323--336.

\bibitem{7605458}
{\sc Mohamed, M., Shrestha, B., and Saxena, N.}
\newblock Smashed: Sniffing and manipulating android sensor data for offensive
  purposes.
\newblock {\em IEEE Transactions on Information Forensics and Security PP}, 99
  (2016), 1--1.

\bibitem{molaylearning}
{\sc Molay, D., Koung, F.-H., and Tam, K.}
\newblock Learning characteristics of smartphone users from accelerometer and
  gyroscope data.

\bibitem{mukherjee2012intrusion}
{\sc Mukherjee, S., and Sharma, N.}
\newblock Intrusion detection using naive bayes classifier with feature
  reduction.
\newblock {\em Procedia Technology 4\/} (2012), 119--128.

\bibitem{murphy2006naive}
{\sc Murphy, K.~P.}
\newblock Naive bayes classifiers.
\newblock {\em University of British Columbia\/} (2006).

\bibitem{Narain:2014:SLK:2627393.2627417}
{\sc Narain, S., Sanatinia, A., and Noubir, G.}
\newblock Single-stroke language-agnostic keylogging using stereo-microphones
  and domain specific machine learning.
\newblock In {\em Proceedings of the 2014 ACM Conference on Security and
  Privacy in Wireless \&\#38; Mobile Networks\/} (New York, NY, USA, 2014),
  WiSec '14, ACM, pp.~201--212.

\bibitem{7113464}
{\sc Nguyen, T.}
\newblock Using unrestricted mobile sensors to infer tapped and traced user
  inputs.
\newblock In {\em Information Technology - New Generations (ITNG), 2015 12th
  International Conference on\/} (April 2015), pp.~151--156.

\bibitem{owusu2012accessory}
{\sc Owusu, E., Han, J., Das, S., Perrig, A., and Zhang, J.}
\newblock Accessory: password inference using accelerometers on smartphones.
\newblock In {\em Proceedings of the Twelfth Workshop on Mobile Computing
  Systems \& Applications\/} (2012), ACM, p.~9.

\bibitem{panda2007network}
{\sc Panda, M., and Patra, M.~R.}
\newblock Network intrusion detection using naive bayes.
\newblock {\em International journal of computer science and network security
  7}, 12 (2007), 258--263.

\bibitem{Park2011}
{\sc Park, B.-W., and Lee, K.~C.}
\newblock {\em The Effect of Users' Characteristics and Experiential Factors on
  the Compulsive Usage of the Smartphone}.
\newblock Springer Berlin Heidelberg, Berlin, Heidelberg, 2011.

\bibitem{peiravian2013machine}
{\sc Peiravian, N., and Zhu, X.}
\newblock Machine learning for android malware detection using permission and
  api calls.
\newblock In {\em Tools with Artificial Intelligence (ICTAI), 2013 IEEE 25th
  International Conference on\/} (2013), IEEE, pp.~300--305.

\bibitem{Petracca:2015:APA:2818000.2818005}
{\sc Petracca, G., Sun, Y., Jaeger, T., and Atamli, A.}
\newblock Audroid: Preventing attacks on audio channels in mobile devices.
\newblock In {\em Proceedings of the 31st Annual Computer Security Applications
  Conference\/} (New York, NY, USA, 2015), ACSAC 2015, ACM, pp.~181--190.

\bibitem{Ping:2015:TIL:2766498.2766511}
{\sc Ping, D., Sun, X., and Mao, B.}
\newblock Textlogger: Inferring longer inputs on touch screen using motion
  sensors.
\newblock In {\em Proceedings of the 8th ACM Conference on Security \& Privacy
  in Wireless and Mobile Networks\/} (New York, NY, USA, 2015), WiSec '15, ACM,
  pp.~24:1--24:12.

\bibitem{pongaliur2008securing}
{\sc Pongaliur, K., Abraham, Z., Liu, A.~X., Xiao, L., and Kempel, L.}
\newblock Securing sensor nodes against side channel attacks.
\newblock In {\em High Assurance Systems Engineering Symposium, 2008. HASE
  2008. 11th IEEE\/} (2008), IEEE, pp.~353--361.

\bibitem{poslad2011ubiquitous}
{\sc Poslad, S.}
\newblock {\em Ubiquitous computing: smart devices, environments and
  interactions}.
\newblock John Wiley \& Sons, 2011.

\bibitem{roesner2012user}
{\sc Roesner, F., Kohno, T., Moshchuk, A., Parno, B., Wang, H.~J., and Cowan,
  C.}
\newblock User-driven access control: Rethinking permission granting in modern
  operating systems.
\newblock In {\em 2012 IEEE Symposium on Security and Privacy\/} (2012), IEEE,
  pp.~224--238.

\bibitem{sahs2012machine}
{\sc Sahs, J., and Khan, L.}
\newblock A machine learning approach to android malware detection.
\newblock In {\em Intelligence and security informatics conference (eisic),
  2012 european\/} (2012), IEEE, pp.~141--147.

\bibitem{schlegel2011soundcomber}
{\sc Schlegel, R., Zhang, K., Zhou, X.-y., Intwala, M., Kapadia, A., and Wang,
  X.}
\newblock Soundcomber: A stealthy and context-aware sound trojan for
  smartphones.
\newblock {\em NDSS 11\/} (2011), 17--33.

\bibitem{schmidt2009static}
{\sc Schmidt, A.-D., Bye, R., Schmidt, H.-G., Clausen, J., Kiraz, O., Yuksel,
  K.~A., Camtepe, S.~A., and Albayrak, S.}
\newblock Static analysis of executables for collaborative malware detection on
  android.
\newblock In {\em Communications, 2009. ICC'09. IEEE International Conference
  on\/} (2009), IEEE, pp.~1--5.

\bibitem{shabtai2012andromaly}
{\sc Shabtai, A., Kanonov, U., Elovici, Y., Glezer, C., and Weiss, Y.}
\newblock “andromaly”: a behavioral malware detection framework for android
  devices.
\newblock {\em Journal of Intelligent Information Systems 38}, 1 (2012),
  161--190.

\bibitem{shen2015input}
{\sc Shen, C., Pei, S., Yang, Z., and Guan, X.}
\newblock Input extraction via motion-sensor behavior analysis on smartphones.
\newblock {\em Computers \& Security 53\/} (2015), 143--155.

\bibitem{Shukla:2014:BYH:2660267.2660360}
{\sc Shukla, D., Kumar, R., Serwadda, A., and Phoha, V.~V.}
\newblock Beware, your hands reveal your secrets!
\newblock In {\em Proceedings of the 2014 ACM SIGSAC Conference on Computer and
  Communications Security\/} (New York, NY, USA, 2014), CCS '14, ACM,
  pp.~904--917.

\bibitem{Simon:2013:PSI:2516760.2516770}
{\sc Simon, L., and Anderson, R.}
\newblock Pin skimmer: Inferring pins through the camera and microphone.
\newblock In {\em Proceedings of the Third ACM Workshop on Security and Privacy
  in Smartphones \&\#38; Mobile Devices\/} (New York, NY, USA, 2013), SPSM '13,
  ACM, pp.~67--78.

\bibitem{smalley2013security}
{\sc Smalley, S., and Craig, R.}
\newblock Security enhanced (se) android: Bringing flexible mac to android.
\newblock In {\em NDSS\/} (2013), vol.~310, pp.~20--38.

\bibitem{son2015rocking}
{\sc Son, Y., Shin, H., Kim, D., Park, Y.-S., Noh, J., Choi, K., Choi, J., Kim,
  Y., et~al.}
\newblock Rocking drones with intentional sound noise on gyroscopic sensors.
\newblock In {\em USENIX Security\/} (2015), pp.~881--896.

\bibitem{Spreitzer:2014:PSE:2666620.2666622}
{\sc Spreitzer, R.}
\newblock Pin skimming: Exploiting the ambient-light sensor in mobile devices.
\newblock In {\em Proceedings of the 4th ACM Workshop on Security and Privacy
  in Smartphones \&\#38; Mobile Devices\/} (New York, NY, USA, 2014), SPSM '14,
  ACM, pp.~51--62.

\bibitem{strikos2007full}
{\sc Strikos, A.~A.}
\newblock A full approach for intrusion detection in wireless sensor networks.
\newblock {\em School of Information and Communication Technology\/} (2007).

\bibitem{6654855}
{\sc Subramanian, V., Uluagac, S., Cam, H., and Beyah, R.}
\newblock Examining the characteristics and implications of sensor side
  channels.
\newblock In {\em Communications (ICC), 2013 IEEE International Conference
  on\/} (June 2013), pp.~2205--2210.

\bibitem{Sun:2014:DIA:2664243.2664245}
{\sc Sun, M., Zheng, M., Lui, J. C.~S., and Jiang, X.}
\newblock Design and implementation of an android host-based intrusion
  prevention system.
\newblock In {\em Proceedings of the 30th Annual Computer Security Applications
  Conference\/} (New York, NY, USA, 2014), ACSAC '14, ACM, pp.~226--235.

\bibitem{tippenhauer2011requirements}
{\sc Tippenhauer, N.~O., P{\"o}pper, C., Rasmussen, K.~B., and Capkun, S.}
\newblock On the requirements for successful gps spoofing attacks.
\newblock In {\em Proceedings of the 18th ACM conference on Computer and
  communications security\/} (2011), ACM, pp.~75--86.

\bibitem{6997498}
{\sc Uluagac, A., Subramanian, V., and Beyah, R.}
\newblock Sensory channel threats to cyber physical systems: A wake-up call.
\newblock In {\em Communications and Network Security (CNS), 2014 IEEE
  Conference on\/} (Oct 2014), pp.~301--309.

\bibitem{Radh1310:Passive}
{\sc Uluagac, S., Radhakrishnan, S.~V., Corbett, C.~L., Baca, A., and Beyah,
  R.}
\newblock A passive technique for fingerprinting wireless devices with
  wired-side observations.
\newblock In {\em 2013 IEEE Conference on Communications and Network Security
  (CNS) (IEEE CNS 2013)\/} (Washington, USA, Oct. 2013), pp.~471--479.

\bibitem{Wang:2015:NHM:2802130.2802132}
{\sc Wang, X., Yang, Y., Zeng, Y., Tang, C., Shi, J., and Xu, K.}
\newblock A novel hybrid mobile malware detection system integrating anomaly
  detection with misuse detection.
\newblock In {\em Proceedings of the 6th International Workshop on Mobile Cloud
  Computing and Services\/} (New York, NY, USA, 2015), MCS '15, ACM,
  pp.~15--22.

\bibitem{wu2012droidmat}
{\sc Wu, D.-J., Mao, C.-H., Wei, T.-E., Lee, H.-M., and Wu, K.-P.}
\newblock Droidmat: Android malware detection through manifest and api calls
  tracing.
\newblock In {\em Information Security (Asia JCIS), 2012 Seventh Asia Joint
  Conference on\/} (2012), IEEE, pp.~62--69.

\bibitem{Wu:2014:DDA:2663761.2664223}
{\sc Wu, W.-C., and Hung, S.-H.}
\newblock Droiddolphin: A dynamic android malware detection framework using big
  data and machine learning.
\newblock In {\em Proceedings of the 2014 Conference on Research in Adaptive
  and Convergent Systems\/} (New York, NY, USA, 2014), RACS '14, ACM,
  pp.~247--252.

\bibitem{Xu:2012:TIU:2185448.2185465}
{\sc Xu, Z., Bai, K., and Zhu, S.}
\newblock Taplogger: Inferring user inputs on smartphone touchscreens using
  on-board motion sensors.
\newblock In {\em Proceedings of the Fifth ACM Conference on Security and
  Privacy in Wireless and Mobile Networks\/} (New York, NY, USA, 2012), WISEC
  '12, ACM, pp.~113--124.

\bibitem{Xu:2015:SPS:2699026.2699114}
{\sc Xu, Z., and Zhu, S.}
\newblock Semadroid: A privacy-aware sensor management framework for
  smartphones.
\newblock In {\em Proceedings of the 5th ACM Conference on Data and Application
  Security and Privacy\/} (New York, NY, USA, 2015), CODASPY '15, ACM,
  pp.~61--72.

\bibitem{ye2000markov}
{\sc Ye, N., et~al.}
\newblock A markov chain model of temporal behavior for anomaly detection.
\newblock In {\em Proceedings of the 2000 IEEE Systems, Man, and Cybernetics
  Information Assurance and Security Workshop\/} (2000), vol.~166, West Point,
  NY, p.~169.

\bibitem{ye2007imds}
{\sc Ye, Y., Wang, D., Li, T., and Ye, D.}
\newblock Imds: Intelligent malware detection system.
\newblock In {\em Proceedings of the 13th ACM SIGKDD international conference
  on Knowledge discovery and data mining\/} (2007), ACM, pp.~1043--1047.

\bibitem{5606038}
{\sc Yu, Y., Wang, J., and Zhou, G.}
\newblock The exploration in the education of professionals in applied internet
  of things engineering.
\newblock In {\em Distance Learning and Education (ICDLE), 2010 4th
  International Conference on\/} (Oct 2010), pp.~74--77.

\bibitem{yu2008framework}
{\sc Yu, Z., and Tsai, J.~J.}
\newblock A framework of machine learning based intrusion detection for
  wireless sensor networks.
\newblock In {\em Sensor Networks, Ubiquitous and Trustworthy Computing, 2008.
  SUTC'08. IEEE International Conference on\/} (2008), IEEE, pp.~272--279.

\bibitem{Zhuang:2009:KAE:1609956.1609959}
{\sc Zhuang, L., Zhou, F., and Tygar, J.~D.}
\newblock Keyboard acoustic emanations revisited.
\newblock {\em ACM Trans. Inf. Syst. Secur. 13}, 1 (Nov. 2009), 3:1--3:26.

\end{thebibliography}
\appendix
\section{Theoretical Foundation}
\subsection{Markov Chain-Based Detection}
%Markov Chain model can be described as a discrete-time stochastic process which denotes a set of random variables and defines how these variables change over time. 
%\begin{itemize}
%\item Probability distribution of the state at time \textit{t+1} depends on the state at time \textit{t} only. There is no effect of previous states leading to the state at time \textit{t} over probability distribution at time \textit{t+1}. Here, the state refers to the overall condition of the stochastic process.
%\item A state transition from previous timestamp (\textit{t}) to next timestamp (\textit{t+1}) is independent of time.
%\end{itemize}
%Markov Chain can be applied to illustrate a series of events where and what state will occur next depends only on the previous state. In our work, a series of events represents user activity and states represent conditions (i.e, values, on/off status) of the sensors in a smart device. We can represent the probabilistic condition of Markov Chain as in equation~\ref{eq1} Equation 1 where \textit{$X_t$} denotes the state at time \textit{t} \cite{keilson2012markov}.

%\begin{equation}
% \begin{split}
% \begin{aligned}
% P(X_{t+1} = x| X_1 = x_1, X_2 = x_2 ..., X_t = x_t) = 
% \\P(X_{t+1} = x| X_t = X_t) , \\ 
% when,\ P(X_1 = x_1, X_2 = x_2 ..., X_t = x_t) > 0
% \end{aligned}
% \end{split}
% \end{equation}

For the Markov Chain detection model, 6thSense observes the changes of condition of a set of sensors as a variable which changes over time. The condition of a sensor indicates whether the sensor value is changing or not from a previous value in time. %Let us assume that \textit{S} denotes a set which represents current conditions of \textit{n} number of sensors. So, \textit{S} can be represented as follows.
%
% \begin{equation}
% \begin{aligned}
% S = \{S_1, S_2, S_3, ... , S_n\}, 
% \\ S_1, S_2, S_3, ... , S_n = 0\ or\ 1
% \end{aligned}
% \end{equation}
%
For a specific time, \textit{t}, 6thSense considers the combination of all the sensors' condition in the smart device as the state of our model. As 6thSense considers change in a sensor's condition as binary output (1 or 0, where 1 denotes that sensor value is changing from previous instance and 0 denotes that sensor value is not changing), the number of total states of in the detection model will be exponents of 2. For example, if we consider the total number of sensors in set \textit{S} is 10, the number of states in our Markov Chain will be $2^{10}$ or 1024 and the states can be represented as a 10 bit binary number where each bit will represent the state of a corresponding sensor. Assume that  \textit{${p_{ij}}$} denotes the probability that the system in a state \textit{j} at time \textit{t+1} given that system is in state \textit{i} at time \textit{t}. If we have \textit{n} number of sensors  and \textit{${m=2^{n}}$} states in our model, 
%
% Markov Chain can be constructed by the following transition probability matrix:
% \begin{equation}
% P=\begin{bmatrix}
% p_{11} & p_{12} & p_{13} & \ldots & \ldots & p_{1m}\\
% p_{21} & p_{22} & p_{23} & \ldots & \ldots & p_{2m}\\
% \ldots & \ldots & \ldots & \ldots & \ldots & \ldots \\
% \ldots & \ldots & \ldots & \ldots & \ldots & \ldots \\
% p_{m1} & p_{m2} & p_{m3} & \ldots & \ldots & p_{mm} \\
% \end{bmatrix}
% \end{equation}
the transition probability matrix of this Markov Chain can be constructed by observing the transitions from one state to another state for a certain time. Assume that 6thSense's states are \textit{${X_0, X_1, \ldots, X_T}$} at a given time \textit{${t= 0, 1, \ldots, T}$}. We can represent the transition probability matrix \cite{ye2000markov} with ${P_{ij}} = \frac{N_{ij}}{N_i}$
%
% \begin{equation}
% {P_{ij}} = \frac{N_{ij}}{N_i},
% \end{equation}
%
with \textit{${N_{ij}}$} = the number of transitions from \textit{${X_t}$} to \textit{${X_{t+1}}$}, where \textit{${X_t}$} in state \textit{i} and \textit{${X_{t+1}}$} in state \textit{j}; \textit{${N_{i}}$} = the number of transitions from \textit{${X_t}$} to \textit{${X_{t+1}}$}, where \textit{${X_t}$} in state \textit{i} and \textit{${X_{t+1}}$} in any other state. The initial probability distribution of this Markov Chain can be as follows \cite{jha2001markov}:
\begin{equation}
Q = \begin{bmatrix}
q_1 & q_2 & q_3 & \ldots & \ldots & q_m
\end{bmatrix}
\end{equation}

where, \textit{${q_m}$} is the probability that the model is in state \textit{m} at time 0. The  probability of observing a sequence of states \textit{${X_1, X_2, \ldots, X_T}$} at a given time \textit{${1, \ldots, T}$} can be computed using the following equation:

\begin{equation}
P(X_1, X_2, \ldots, X_T) = {q_{x1}} \prod_{2}^{T} {P_{{X_{t-1}}X{t}}}
\end{equation}

As described earlier in Section 5, for 6thSense, we use a modified version of the general Markov Chain model. Instead of predicting the next state, 6thSense determines the probability of a transition occurring between two states at a given time. %We train the Markov Chain model with a training dataset collected from real users (50 users total) and build the transition matrix accordingly. Then, 6thSense determines the condtions of sensors for time \textit{t} and \textit{t+1}. Let us assume \textit{a} and \textit{b} are a sensor's states in time \textit{t} and \textit{t+1}. 6thSense looks up for the probability of transition from state \textit{a} to \textit{b}, which can be found by looking up in the transition matrix and calculating \textit{P(a,b)}. As the training dataset consisted of sensor data from benign activities, we can assume that if transition from state \textit{a} to \textit{b} is malicious, the calculated probability from the transition matrix will be zero.

\subsection{Naive Bayes Based Detection}
Naive Bayes model is a simple probability estimation method which is based on Bayes' method. The main assumption of the Naive Bayes detection is the 
%\begin{itemize}
%\item 
presence of a particular particular  sensor condition  in  a  task/activity %feature in a class
 has no influence over the presence of any other feature on that particular event.
%\end{itemize}

Assume \textit{${p(x_1,x_2)}$} is the general probability distribution of two events \textit{${x_1, x_2}$}. Using the Bayes rule, we can have $p(x_1,x_2) = p(x_1|x_2)p(x_2)$ 
%
%the following equation: 
%\begin{equation}\label{eq2}
%p(x_1,x_2) = p(x_1|x_2)p(x_2)
%\end{equation}
where, \textit{${p(x_1|x_2)}$} = Probability of event \textit{${x_1}$} given that event \textit{${x_2}$} will happen. Now, with \textit{c}, we can rewrite this formula as $p(x_1,x_2|c) = p(x_1|x_2,c)p(x_2|c)$. 
%
% Equation~\ref{eq2} as follows: 
% \begin{equation}\label{eq3}
% p(x_1,x_2|c) = p(x_1|x_2,c)p(x_2|c)
% \end{equation}
If \textit{c} is sufficient enough to determine the probability of event \textit{${x_1}$}, we can state that there is conditional independence between \textit{${x_1}$} and \textit{${x_2}$}~\cite{panda2007network}. So, we can rewrite the first part as \textit{${p(x_1|x_2,c) = p(x_1|c)}$}, which then modifies the formula as follows: 
\begin{equation}\label{eq4}
p(x_1,x_2|c) = p(x_1|c)p(x_2|c)
\end{equation}

% In 6thSense, we consider users' activity as a combination of \textit{n} number of sensors. Assume, X is a set which represents current conditions of \textit{n} number of sensors. We consider that conditions of sensors are conditionally independent, which means a change in one sensor's working condition (i.e., on/off state) has no effect over a change in another sensor's working condition. As  explained earlier, the probability of executing a task depends on the conditions of a specific set of sensors. So, in summary, although one sensor's condition does not control another sensor's condition, the overall probability depends on all the sensors' conditions. For example, if a person is walking with his smartphone in his hand, the motion sensors (accelerometer and gyroscope) will change. However, this change will not force the light sensor or the proximity sensor to change its condition. Thus, sensors in a smartphone change their conditions independently, but execute a task together. From Equation~\ref{eq4}, we can have a generalized formula for this context-aware model \cite{mukherjee2012intrusion}: 
6thSense considers users' activity as a combination of \textit{n} number of sensors. Assume X is a set which represents current conditions of \textit{n} number of sensors. We consider that conditions of sensors are conditionally independent (See Section 4.2), which means a change in one sensor's working condition (i.e., on/off states) has no effect over a change in another sensor's working condition. As explained earlier, the probability of executing a task depends on the conditions of a specific set of sensors. So, in summary, although one sensors' condition does not control another sensor's condition, overall the probability of executing a specific task depends on all the sensors' conditions. As an example, if a person is walking with his smartphone in his hand, the motion sensors (accelerometer and gyroscope) will change. However, this change will not force the light sensor or the proximity sensor to change its condition. Thus, sensors in a smartphone change their conditions independently, but execute a task together. We can have a generalized model for this context-aware detection \cite{mukherjee2012intrusion} as follows: 
\begin{equation}
p(X|c)=\prod_{i=1}^{n} p(X_i|c)
\end{equation}

In 6thSense's context-aware activity-oriented detection model, we have a set of training data for users' activities. Assume that  \textit{B} represents a set which denotes \textit{m} numbers of user activities. We can determine the probability of a dataset $X$ to be classified as a user activity using the following equation: 

\begin{equation}
P(B_i|X) = \frac{P(X|B_i)P(B_i)}{P(X)}, 
\end{equation}

where $i = 1, 2$, \ldots , $m$. As the sum of all the conditional probabilities for $X$ will be 1, we can have the following equation,  which then will lead to Equation~\ref{eq5}~\cite{kruegel2003bayesian}: 

\begin{equation}
\sum_{i=1}^m P(B_i|X) = 1. 
\end{equation}

\begin{equation}\label{eq5}
P(B_i|X) = \frac{P(X|B_i)P(B_i)}{\sum_{i=1}^m P(X|B_i) P(B_i)}.
\end{equation}

This calculated conditional probability then is used to determine the benign user activity or malicious attacks in 6thSense. In this way, 6thSense computes the probability of occurring an activity over a certain period of time. 

\par 6thSense divides the sensor data into smaller time values (1 second) and calculates the probability of each instance to infer the user activity. The calculated probability of each second data is then used in the expected value to calculate the total probability. As such, the probability of the first instance is \textit{${p_1}$} with a value of \textit{${a_1}$}, the probability of the second instance is \textit{${p_2}$} with a value of \textit{${a_2}$} and so on up to the value \textit{${a_n}$}. Then,  the expected value can be calculated by the following formula:

\begin{equation}
E[N]= \frac{a_1p_1+a_2p_2+a_3p_3+\ldots \\ \ldots+a_np_n}{a_1+a_2+\ldots \ \ldots+ a_n}.
\end{equation}

As all the values of \textit{${a_1}$}, \textit{${a_2}$}, ... ..., \textit{${a_n}$} are equally likely, this expected value becomes a simple average of the cumulative probability of each instance. In this way, 6thSense infers the user activity by setting up a configurable threshold value and checking whether the calculated value is higher than the threshold or not. If it is lower than the threshold value, 6thSense concludes that the malicious activity is occurring in the smart device.

%\section{Figures}
%\begin{figure}[h!]
%  \centering
%  \framebox{  
%    % \includegraphics[width=\textwidth,height=8cm]{new3-1.png}
%    \includegraphics[height=6cm, width=7cm]{android_sensor_framework_2}
%    }
%      \caption{Android Sensor Management Architecture}
%      \label{overviewnew}
%\end{figure}
%\begin{figure}[h!]
%  \centering
%%  \framebox{  
%    % \includegraphics[width=\textwidth,height=8cm]{new3-1.png}
%    \includegraphics[height=6cm, width=7cm]{Accuracy_Fscore_Naive}
%    }
%      \caption{Naive Bayes model results}
%      \label{overviewnew4}
%\end{figure}
%\begin{figure}[h!]
%  \centering
%  \framebox{  
%    % \includegraphics[width=\textwidth,height=8cm]{new3-1.png}
%    \includegraphics[height=5cm, width=9cm]{Framework_phases_2}
%    }
%%      \caption{Framework phases}
% %     \label{overviewnew2}
%\end{figure}

\end{document}